\documentclass[letterpaper,twocolumn,10pt]{article}
\usepackage{usenix-2020-09}
\pdfoutput=1
\usepackage{cite}
\usepackage{tikz}
\usepackage{amsmath}

\usepackage{tabularx}
\usepackage{multirow}
\usepackage{subfig}
\usepackage{subfloat}
\usepackage{sidecap}
\usepackage{graphicx}
\usepackage{textcomp}
\usepackage{booktabs} 
\usepackage{soul}
\usepackage{pifont}
\usepackage{amsmath}
\usepackage{algorithm}
\usepackage{algpseudocode}
\usepackage{multibib}
\usepackage{lipsum}
\usepackage{subfloat}
\usepackage{comment}
\usepackage{titlesec}
\usepackage{multirow}
\usepackage{amssymb}

\newcommand\revision[1]{\textcolor{black}{#1}}
\usepackage[utf8]{inputenc}

\usepackage{filecontents}
\newcommand{\sys}{REMARK-LLM}
\newcommand{\baseend}{AWT}
\newcommand{\baseinfer}{KGW}
\newcommand{\baserule}{CATER}
\newcommand{\baseinferrb}{EXP}

\newcommand{\cmark}{\ding{51}}%
\newcommand{\xmark}{\ding{55}}%
\graphicspath{ {./figs/} }
\begin{document}
%-------------------------------------------------------------------------------

%don't want date printed
\date{}

%\title{\sys: A \emph{R}obust and \emph{E}fficient Water\emph{mark}ing Framework for \\Generative Large Language Models}
\title{\sys: A \emph{R}obust and \emph{E}fficient Water\emph{mark}ing Framework for \\Generative Large Language Models}
\author{Ruisi Zhang, Shehzeen Samarah Hussain, Paarth Neekhara, Farinaz Koushanfar \\
\textit{University of California, San Diego}}

\maketitle
%-------------------------------------------------------------------------------
\begin{abstract}
We present \sys, a novel efficient, and robust watermarking framework designed for texts generated by large language models (LLMs). Synthesizing human-like content using LLMs necessitates vast computational resources and extensive datasets, encapsulating critical intellectual property (IP). However, the generated content is prone to malicious exploitation, including spamming and plagiarism. 
To address the challenges, \sys{} proposes three new components: (i) a learning-based message encoding module to infuse binary signatures into LLM-generated texts; (ii) a reparameterization module to transform the dense distributions from the message encoding to the sparse distribution of the watermarked textual tokens; (iii) a decoding module dedicated for signature extraction; 
%Furthermore, we introduce an optimized beam search algorithm to guarantee the coherence and consistency of the generated content. 
\revision{Besides, we introduce an optimized beam search algorithm to generate content with coherence and consistency. }
\sys{} is rigorously trained to encourage the preservation of semantic integrity in watermarked content, while ensuring effective watermark retrieval. Extensive evaluations on multiple unseen datasets highlight \sys's proficiency and transferability in inserting 2$\times$ more signature bits into the same texts when compared to prior art, all while maintaining semantic integrity. Furthermore, \sys{} exhibits better resilience against a spectrum of watermark detection and removal attacks.
\end{abstract}

\section{Introduction}
\label{sec:intro}
% What is the problem;  in generating human-like text responses to questions
Recent advancements in the development of large language models (LLMs) such as ChatGPT~\cite{schulman2022chatgpt}, LLaMA~\cite{touvron2023llama}, and GPT-4~\cite{openai2023gpt} indicate a paradigm shift in human-computer dialogue interactions. 
These AI-powered systems have the capacity to generate human-like text responses and are integrated into various aspects of our daily lives. Training LLMs~\cite{bommasani2021opportunities,narayan2022can,orr2022data} requires substantial computational resources and extensive datasets, both of which are critical components of valuable intellectual property (IP).
Concurrently, the increasing capabilities of these foundational models present potential risks in the form of malicious uses, including spam and plagiarism.
% Concurrently, the growing potential of these foundation models~\cite{bommasani2021opportunities,narayan2022can,orr2022data} poses latent threats in malevolent exploitations such as spamming and plagiarism. 
Thus, there is a need to devise mechanisms to claim ownership of LLM-generated text and trace the distribution of the generated content.

%proprietors of these LLMs need to devise mechanisms to claim their ownership or enable trace the distribution of their generated contents.

%Consequently, owners of these foundational models need techniques to trace the usage of the content they generate. The outputs produced by these models often encompass valuable intellectual property, which necessitates stringent safeguarding measures. 
% and are in need of protection.  
%Therefore, establishing a clear trail of accountability is essential for regulating the use of LLMs.

%What has been done
Watermarking offers a promising solution to tackle two persistent issues: asserting ownership of generated output and tracing the source of content. 
% of claiming ownership of generated output and content tracing. 
By embedding watermark signatures into the outputs of LLMs, model proprietors can effectively monitor their content utilizations and validate their ownership. 
\revision{As such, the system can be applied to detect plagiarism to maintain academic integrity~\cite{lancaster2023artificial}, protect the copyrights of LLM owners~\cite{li2023protecting}, and track the distribution of the potential misinformation generated by LLMs~\cite{megias2022architecture}.}
Existing literature on text watermarking can be classified into three categories~\cite{tang2023science}: (1) rule-based watermarking~\cite{kim2003text,munyer2023deeptextmark}, (2) inference-time watermarking~\cite{kirchenbauer2023watermark,liu2023private}, and (3) neural-based watermarking~\cite{abdelnabi21oakland}. 
The rule-based watermarking replaces synonym~\cite{keskisarkka2012automatic} or transforms syntactic structures~\cite{chalmers1992syntactic} in the paragraph to insert as watermarks. 
% Those 
Such manually designed features 
% made 
make the inserted signatures statistically removable through word distribution or syntactical analysis.
The inference-time watermarking~\cite{kirchenbauer2023watermark} splits vocabulary into green/red lists and restricts the LLM decoding to predict the next tokens from the green list. While the inserted watermarks are robust against attacks, the decoding strategy drastically distorts
the semantic similarity between the watermarked and original LLM outputs. The neural-based approach~\cite{abdelnabi21oakland} leverages an end-to-end learning technique to integrate the binary watermarking signatures into the LLM-generated texts while maintaining semantic coherence.  However, the maximum encodable signature length per token segment is limited compared with the rule-based and inference-time frameworks, thus hindering the practical usage of this approach.

Generally speaking, watermarking text data presents several challenges. First, text data exhibits a pronounced sparsity compared with other modalities, such as images and audio. For instance, a 256-pixel image offers approximately 65k feasible positions for watermark insertion~\cite{neekhara2022facesigns}, whereas the maximum token limit in GPT-4~\cite{gpt-4-link} is 8.2k. Besides, text data is fragile in that subtle alterations may obfuscate or compromise the semiotic fidelity~\cite{zhao2023protecting}, whereas minor perturbations in images can remain imperceptible. Note that all inserted watermarks, including the ones applied to LLM-generated text, should be resilient to potential watermark removal and detection attacks from end users~\cite{kirchenbauer2023reliability}.

% It hinders the neural-based approach to be applied in LLM-generated texts. 

%\textcolor{red}{SH: There should be some lines about the motivation for this work. Why do we want to embed watermarks in text sequences?}
 
%\textcolor{red}{Paragraph 2: What has been done; }
%Many prior works try to detect malicious activities from end-users by embedding invisible watermarks into LLM-generated texts~\cite{abdelnabi21oakland,kirchenbauer2023watermark} and help claim intellectual property (IP) ownership. However, those watermarking approaches can modify the text semantics and are not resilient to well-designed potential watermark removal attacks~\cite{tang2023science}.

%  What are the challenges and limitations of current practice

%Nevertheless, inserting watermarks into text data is nontrivial and challenging for the following reasons. First of all, texts are much sparser compared to images. Given a 256-pixel image, there are 65k candidate positions to insert watermarks, whereas the maximum number of tokens in GPT-4 is 8.2k~\cite{gpt-4-link}. Secondly, watermarking should not alter the semantics of the original LLM-generated texts and guarantee that the texts are readable. Small changes in the images will not affect the users' perception, whereas small errors in the texts will
% crush 

%  What is it that we are doing
This paper proposes \sys{}, a new robust and efficient watermarking framework to insert watermarks into LLM-generated texts without compromising the semantic integrity.
\sys{} composed of three key modules, namely, message encoding, reparameterization, and message decoding. The message encoding module encodes the LLM-generated texts and their corresponding signatures into latent feature space. Their feature representations are added and yield the watermarked distribution over the vocabulary.  Sequentially, the reparameterization component exploits Gumbel-Softmax methodology~\cite{jang2016categorical} to transform the watermarked distribution to the sparse distribution of the watermarked textual tokens. Next, the message decoding module extracts watermarking signatures by leveraging a transformer to predict the inserted messages. \sys{} enhances its robustness by incorporating malicious transformations during training, including text addition, deletion, and substitution over the transformed textual token distribution into the message decoding phase.

The three modules are trained end-to-end, targeting to (1) preserve the semantic fidelity by minimizing a semantic loss between the original LLM-generated and watermarked texts, (2) ensure watermark extraction by minimizing a message recovery loss between the inserted and extracted watermarking signatures from the watermarked texts, and  
 (3) enhance robustness by extracting watermarking signatures from the malicious transformations.

% Then, the reparameterization module mimics the categorical distribution of the watermarked token distributions by applying Gumbel softmax~\cite{jang2016categorical}. The mimicked categorical distribution is close to the one-hot encoding of the watermarked text tokens and is mapped into its embedding space via a linear layer. Then, message decoding decodes the watermark signatures from the watermarked embeddings using a small decoding transformer model. The three modules are trained in an end-to-end manner by minimizing two losses simultaneously: the first one is the message loss between the input signatures and the decoded messages, and the second one is the semantic loss between the original texts and the reparameterized watermark embeddings. 

With the trained \sys, the LLM proprietor leverages the message encoding module to embed binary signatures into the LLM-generated texts and obtain a watermarked distribution. An optimized beam search algorithm subsequently translates the output of this module's distribution into watermarked texts, ensuring linguistic coherence, unwavering semantic fidelity, and the successful extraction of signatures. Next, the watermarked texts are disseminated to end-users as coherent responses. 
The watermark existence can be verified by extracting  
the inserted signatures using the message decoding module. It compares the extracted messages with the inserted signatures to determine if the LLM generates the texts. 

%The watermarked texts are first converted to one-hot distribution over the vocabulary space, then extracted the messages from \sys's decoding module via a comparison with the inserted signatures. 

%It ensures the converted distributions are closer to the one-hot encoding distributions from the texts and differentiable during end-to-end training. 

%The Message Encoding inserts invisible watermarks into the input texts.  Reparameterization converts the watermarked token distribution from the Seq2Seq model closer to the categorical distribution and maps the parameterized distribution into their embedding representation again. \textbf{Message Decoding} employs a decoder to extract the secret messages from the embedding.  The three modules are trained in an end-to-end manner, aiming minimizing the semantic loss between the watermarked texts and the original texts and the message extraction loss between the embeded message and the extracted message. 

%The watermark insertion employs an optimized beam search algorithm to decode the watermarked texts from the token distributions in \textbf{Message Encoding}. It ensures the watermarked messages are extractable while maintaining the readabilty and semantics from the generated texts.
%To extract watermarks, we apply the \textbf{Message Decoding} modules trained in the end-to-end training stage to extract the watermarks. 
 
In 
% brief, 
summary, our contributions are  as follows:
\begin{itemize}
\item  We introduce \sys, a robust and efficient watermarking framework tailored for LLMs that maintains the semantic integrity of watermarked text while exhibiting resilience against potential watermark detection and removal attacks. 
%\item  We propose a modular three-phase watermark approach based on message encoding, reparameterization, and decoding. 
%To ensure textual coherence and consistency, we optimize the beam search algorithm to better decode the signed text from the sparse token distribution. 
\item \revision{We devise the watermarking framework with novelties lie in (i) a pre-trained sequence-to-sequence backbone module to significantly improve the transferability and capacity of the watermarking framework; (ii) an optimizing beam search module for balanced readability and extraction accuracy; (iii) incorporate potential malicious transformation into the training for improved robustness against watermark removal attacks.}
\item We validate the effectiveness and robustness of \sys{} by conducting extensive evaluations on multiple datasets: 
(i) \sys{} can successfully embed 2$\times$ more signatures into LLM-generated text compared to prior art AWT~\cite{abdelnabi21oakland} within 1.5 seconds; (ii) \sys{} maintains the LLM-generated texts' semantics with an average of 0.90 BERT score and exhibits transferability to watermark natural language from unseen sources without extra fine-tuning; 
(iii) \sys{} is resilient under various watermark detection and removal attacks and maintains an average AUC of 0.85.  

%(i) the trained \sys{} is agnostic to texts generated from different LLM architectures, and maintains over 90\% watermark extraction rates with high semantic preservation; (ii) \item (iii) 

%Our analysis confirms \sys{}'s efficacy in handling texts from various LLMs, with over a 90% watermark extraction rate and high semantic preservation. It outperforms predecessors like AWT, embedding twice as many signatures. Notably, \sys{} preserves over 80% extraction efficiency against diverse watermark removal techniques, from editing to re-watermarking

% extraction efficiency against diverse watermark removal techniques, from editing to re-watermarking."

%     \item \sys{} trains the message encoding and decoding modules, as well as a reparameterization component connecting two modules, in an end-to-end manner to insert watermark signatures into the LLM-generated texts without modifying their semantic information. 

%     \item In watermark insertion, \sys{} employs an optimized beam search algorithm to decode watermarked texts from the message encoding module, preserving both readability and high watermark extraction rates. 

 %    \item Experiments on various datasets have demonstrated that \sys{} can successfully embed up to 16-bit binary watermark signatures within sentences composed of 80 tokens without compromising semantic fidelity.

%    \item By including a variety of adversarial attack examples during training, \sys{} exhibits robustness against watermark removal attacks. 
\end{itemize}

\textbf{Paper Organization:} The rest of the paper is organized as follows:
Section~\ref{sec:background} provides the background and related work on text watermarking.
Section~\ref{sec:goals} describes the problem formulation, including the watermarking objectives, challenges, and potential threats.
Section~\ref{sec:method} introduces the proposed watermarking scheme \sys, by detailing the watermarking architecture, as well as the signature insertion and extraction at the inference time. 
Section~\ref{sec:exp} details the extensive experiments on different datasets and comparisons with existing watermark schemes regarding effectiveness, efficiency, and robustness.
%baselines to demonstrate \sys's effectiveness.
Finally, Section~\ref{sec:conclusion} summarizes the work.

\section{Background and Related Work}
\label{sec:background}
In this section, we first introduce the background and related work for LLM watermarking. Then, we compare the capabilities of those watermarking techniques. 

\subsection{LLM Watermarking}
%\Watermarking LLM-generated texts can be approached in two directions.  The first direction involves integrating post-hoc watermarks onto LLM-generated texts. The second direction is embedding secret signals into the LLM decoding to defend against model extraction attacks. It will trigger  %If the adversary trains an extracted model from the LLM outputs, these signals, when activated with probe inputs, will induce variant behaviors in the extracted model, serving as watermarks. 

%\\subsubsection{Text Watermarking}
Adding post-hoc watermarks in LLM-generated texts can be methodologically categorized into~\cite{tang2023science}: (1) Rule-based watermarking, (2) Inference-time Watermarking, and (3) Neural-based Watermarking. 

\noindent\textbf{Rule-based watermarking}  This approach integrates watermarks into LLM-generated texts by manipulating linguistic features~\cite{he2022cater,yoo2023robust}, altering lexical properties~\cite{he2022protecting}, and substituting synonyms~\cite{munyer2023deeptextmark,yang2023watermarking}.
The rule-based watermarking approach aims to insert the synonym replacement or syntactic transformations as watermarks while ensuring the overall semantics are not distorted. 
%However, they embed watermarks at a superficial text level, susceptible to removal attacks like text rephrasing or editing attacks.

\noindent\textbf{Inference-time Watermarking} Inference-time watermarking~\cite{kirchenbauer2023watermark,kirchenbauer2023reliability} approach inserts signatures at the LLM decoding stage. This approach divides the vocabulary into red/green lists and only allows LLM to decode tokens from the green list.  Some follow-up works~\cite{christ2023undetectable,zhao2023provable} proposed different red/green list splitting algorithms or sampling algorithms from the green list probabilistic distribution to enhance the explainability and robustness of the inference-time watermarking. 

\noindent\textbf{Neural-based Watermarking} A neural based approach~\cite{abdelnabi21oakland} encodes LLM-generated texts and associated message signatures through an end-to-end machine learning paradigm. It leverages a data-hiding network to infuse the watermark signatures into the LLM-generated texts and a data-revealing network to decode the signature from the watermark text. 
This facilitates the signatures to be integrated into the feature space of the watermarked text without compromising the semantic fidelity. However, a notable limitation of current \revision{state-of-the-art} neural-based approach AWT~\cite{abdelnabi21oakland} is the limited embeddable signature length capacity~\cite{tang2023science}.

%\subsubsection{Model Watermarking}
%\Model watermarking was first introduced in image classification tasks~\cite{qiao2023novel,zhang2021deep,chen2019deepmarks,darvish2019deepsigns}. These frameworks embed unique signatures into models by modifying the model weight parameters' probability distribution or introducing backdoor triggers during the training phase. 
%\In the natural language processing (NLP) field, the model watermarking~\cite{charette2022cosine,zhao2023protecting} integrates secret signals as the probability vector during token decoding at the inference phrase. This approach primarily defends against model extraction attacks wherein adversaries leverage batch responses from the target LLM as training sets to train a downsized model.
%\LLM proprietors can verify whether the model is an extraction from the LLM  by posing a designated probing input. A matching output to the secret signal affirms the extraction origin of the training data.

\subsection{Comparison}

An ideal text watermarking framework should adhere to the following three criteria: 

\noindent\textbf{Criteria 1 Effectiveness}: The inserted watermark signatures can be seamlessly extracted. % without compromising content quality. This entails that signature insertion not only preserves the original semantics but also ensures that the text coherence and consistency remain undistorted.

\noindent\revision{\textbf{Criteria 2 Fidelity}: The watermarked content quality shall not be compromised. This entails that signature insertion not only preserves the original semantics but also ensures that the text coherence and consistency remain undistorted.}

\noindent\textbf{Criteria 3 Efficiency}: The watermark insertion and extraction are efficient. This includes both minimal computation and time overheads to ensure rapid IP insertion/verification without excessive computational resources.

\noindent\textbf{Criteria 4 Robustness}: Resilience against potential threats is crucial to help LLM proprietors verify IP and trace data sources. Therefore, the signatures shall remain extractable under watermark detection and removal attacks.

\noindent\revision{\textbf{Criteria 5 Undetectability}: The watermarks are invisible upon inspection. As a result, the adversary cannot detect whether a given text is watermarked. }

We systematically evaluate the capabilities of the watermarking frameworks previously against the proposed criteria in Table~\ref{tab:compare_scheme}. \revision{The rule-based watermarking, like \baserule~\cite{he2022cater}, demonstrates effectiveness and efficiency by inserting watermarks in the linguistic attributes of the texts. However, the adversary may exploit syntactic transformations or synonym replacements to remove manually designed watermarking signatures from \baserule~\cite{he2022cater}.}

\revision{The inference-time watermarking achieves resilience by embedding watermarks during each token decoding. However, \baseinfer~\cite{kirchenbauer2023watermark} introduces semantic discrepancies between the watermarked and original texts, undermining the LLM's fidelity. While \baseinferrb~\cite{kuditipudi2023robust} tries to improve the semantic preservation, the efficiency is compromised compared to \baseinfer~\cite{kirchenbauer2023watermark}.} 

%The inference-time watermarking showcased resilience by embedding watermarks during each token decoding. However, there are discrepancies between the watermarked and original texts, undermining the LLM's effectiveness. The required computational time for watermarking insertion also becomes larger when integrating watermarks into LLMs at the inference stage. 

\revision{The neural-based approach like \baseend~\cite{abdelnabi21oakland} leverages machine learning algorithms to embed watermarks into the LLM-generated texts without tampering with textual semantics. It achieves efficiency and robustness by embedding the watermarks through text feature space via lightweight language models.   }

To take the best properties of the neural-based watermarking, we devise \sys{} as a pioneering LLM-generated text watermarking methodology. It can embed up to 2$\times$ longer signature sequences into the same contents compared with the best prior art~\cite{abdelnabi21oakland} without compromising the textual semantics and coherence.
%\noindent\textbf{Inference-Time Methodology:} While showcasing resilience, this method embeds watermarks during each token decoding instance. A pitfall is that the modified text exhibits discrepancies from the original LLM-generated outputs, undermining the LLM's optimal performance. The computational overhead, given the watermark's integration during inference, is notably more pronounced than its counterparts.

%\noindent\textbf{Neural-Based Methodology:} This paradigm, harnessing the prowess of machine learning algorithms, embeds watermarks without tampering with the textual semantics, thereby preserving fluency. Its adeptness in watermark insertion, facilitated by lightweight machine-learning algorithms, is superior to the inference-time approach, and it stands robust against myriad adversarial incursions.

%Therefore, following this path, we present \sys{} as an innovative approach that can embed up to 4$\times$ more message length into the same text length compared with the state-of-the-art baseline AWT~\cite{abdelnabi21oakland}. \sys{} does not compromise the text quality while maintaining high watermark extraction rates.

\begin{table}[h]
\centering
 \resizebox{\columnwidth}{!}{
\begin{tabular}{cccccc}
\toprule
Method & Effectiveness & Fidelity & Efficiency & Robustness  & Undetectability \\ \midrule
Rule-based~\cite{he2022cater,yoo2023robust} & \cmark & \cmark & \cmark & \xmark  & \cmark \\  
Inference-time~\cite{kirchenbauer2023watermark,kuditipudi2023robust} & \cmark & \xmark & \xmark &  \cmark  & \cmark\\  
Neural-based~\cite{abdelnabi21oakland} &  \cmark & \cmark &  \cmark  &   \cmark  & \cmark \\  
%Model-based~\cite{charette2022cosine,zhao2023protecting} & \xmark  &  \xmark  &  \cmark \\  
\bottomrule
\end{tabular}
}
\caption{Comparison of post-hoc LLM-generated text watermarking schemes.\label{tab:compare_scheme}}
\vspace{-20pt}
\end{table}

\section{Problem Formulation}
\label{sec:goals}
In this section, we first introduce the watermarking objectives for the LLM-generated texts. Next, we discuss the various challenges of incorporating watermarks into the content and define the threat models that impact watermarked text.
% Then, we introduce several challenges when performing the watermark insertion and the threat models the watermarked texts face.

\subsection{Watermarking Objective}~\label{subsec:goal}

\begin{figure}[!ht]
    \centering
    \includegraphics[width= \columnwidth]{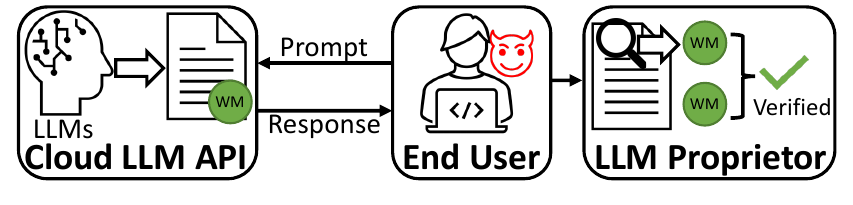}
    \caption{LLM-generated text watermarking scenario. The local user sends prompts to the remote LLM cloud API, and the API watermarks (WM) the responded texts before sending them back to users. LLM proprietor claims ownership by using the message decoding module to decode the signatures and compare them with inserted watermarks. }
    \label{fig:scenario}
    \vspace{-10pt}
\end{figure}

Given the increasing popularity of LLMs
% ChatGPT and GPT-4 
and the heightened prominence of machine-generated media, the dangers associated with the content they produce have become more significant. Content generated by LLMs may inadvertently be employed to create counterfeit essays or inundate the internet with spam responses, thereby posing a threat to the credibility of online content.
Recognizing this challenge, \sys{} equips LLM proprietors with robust IP tracing toolsets as shown in Figure~\ref{fig:scenario}. The local users submit prompt requests to a remote cloud-hosted LLM API to obtain responses. The LLM inserts watermarks into their generated text response before sending it to local users. An LLM proprietor can trace malicious usages online and claim ownership by applying message decoding modules to extract the signatures. LLM proprietors can claim ownership and prove whether texts are machine-generated using \sys{} by comparing the inserted and extracted signatures.

% LLM-generated texts can be unintendedly used to fake essays or spam responses on the Internet, jeopardizing online content's integrity.

\revision{
\textbf{Applications:} The watermarking system has wide applications with the emerging popularity of large language models: (i) Education~\cite{lancaster2023artificial}: Watermarking can help teachers and professors identify whether student homework submissions, like essays or research papers, are AI-generated. This maintains academic integrity and ensures students engage in original thinking and writing. (ii) Copyright Protection~\cite{li2023protecting}: Watermarking can help detect if humans or AI write the given article. It protects the LLM owners' copyrights because the article publishers could make profits from the AI-generated content without proper acknowledgment.
%It not only maintains the trust of readers who may want to know the source of the information but also protects the copyrights of the LLM owners. 
(iii) Misinformation Monitor~\cite{megias2022architecture}: Watermarking can be used by social media platforms to detect and label watermarked AI-generated content automatically. It helps to combat the potential spread of misinformation or inauthentic content.
}

\subsection{Challenges}
When compared to other methods of watermarking various media~\cite{qiao2023novel,zhang2021deep,chen2019deepmarks,darvish2019deepsigns}, the process of embedding watermarks into text data presents distinct challenges, as outlined below.
% Compared with other media watermarking modalities, inserting watermarks into text data exhibits several unique challenges as follows:  

\noindent\textbf{Challenges 1 Sparsity}: 
A given image with size 256$\times$256 provides over 65,000 potential pixel positions for embedding watermarks~\cite{neekhara2022facesigns}. This granularity ensures notable flexibility in watermark accommodation. However, the maximum token length LLMs like GPT-4 can generate for text data is 8.2k~\cite{gpt-4-link}.
This significantly constrains the potential embedding locations and demands a more sophisticated-designed watermarking approach.

\noindent\textbf{Challenges 2 Sensitivity}: Text data exhibits a heightened sensitivity to alterations~\cite{zhao2023protecting}. Minor image pixel adjustments often remain imperceptible, ensuring the image's aesthetic after watermark embedding. Text data, on the other hand, minor changes can distort the intended meanings and make the text incoherent or misleading.  

\noindent\textbf{Challenges 3 Vulnerability}: If an adversary suspects or detects watermarking, they might attempt to remove or alter the inserted signatures by rephrasing or editing attacks~\cite{kirchenbauer2023reliability}. 

%it. Those watermarked texts are suspected to be removed by paraphrasing or editing attacks performed by the adversaries. 

\subsection{Threat Model}~\label{subsec:threat}

\noindent\textbf{Adversary's Capacity}
We assume the adversary is an end-user of the LLM cloud service, where he has black access to the API. However, he does not have access to the trained watermarking models and the original LLM-generated outputs. 
The adversary attempts to exploit the LLM-generated content for malicious usage without being traced. Therefore, he performs attacks to detect and remove the signatures within the watermarked contents without distorting their semantics. 

\revision{This threat model setting is consistent with prior work \baseend~\cite{abdelnabi21oakland} and \baseinferrb~\cite{kuditipudi2023robust} that assume: (i) the watermarking framework is kept private, where adversaries as end-users do not have control over \sys's weights and/or hyperparameters; (2) the adversarial attacks do not greatly compromise the generated output quality, accuracy, and readability. }

%The adversary attempts to steal the IP of the LLM-generated texts by falsely claiming the ownership of the LLM-generated text. The adversary may additionally perform attacks to 
%remove the inserted message signatures that establish LLM-owner's IP. We consider the adversary to have the following capacities:

%\begin{itemize}
%    \item The adversary can randomly edit the sentence to remove the watermark. However, the paragraphs after the edition should preserve semantics for the adversary to use.
%    \item  The adversary can apply other natural language processing(NLP) rephrase models to help rewrite the paragraphs to ensure semantics preservation while changing the syntactic of the texts.
%\end{itemize}

%The objective of \sys{} is to provide content provenance and authentication to prevent the illegal usage of LLM-generated texts, like essay plagiarism or spam reports. 
% Those adversaries aim to batch-generate high-quality texts without extra human efforts in the loop. 
%We do not consider the scenario where the end-users rewrite the paragraph with the contents but exactly different paragraph organizations because the intelligence of the end-users is included in the texts as well. 

\noindent\textbf{Potential Attacks}
%\sys's primary objective is to ensure content authenticity, but it faces threats from adversaries who employ various strategies to remove these watermarks. 
The adversary performs detection attacks to inspect if the LLM-generated contents have watermark insertion or not. 
If the adversary suspects or detects watermarks, he performs attacks to remove inserted signatures by directly manipulating the textual content or leveraging sophisticated NLP models for rephrasing. 

%Such attacks, while simple in concept, can be quite effective in removing embedded signatures. 

\noindent $\bullet$ 
 \textit{\textbf{Attack 1 Watermark Detection Attack}}: The adversary uses statistical analysis or machine learning models to detect whether the texts are watermarked. 

\noindent $\bullet$ 
 \textit{\textbf{Attack 2 Text Edit Attack}}: The adversary doesn't have prior linguistic knowledge. By randomly deleting, adding, or substituting words within the content, he attempts to destroy the watermark while preserving the overall meanings.

\noindent $\bullet$ 
 \textit{\textbf{Attack 3 Text Rephrase Attack}}: The adversary can exploit open-source NLP models, such as T5, to remove watermarks. By feeding the content into these models, the adversary aims to generate a rephrased version of the original texts to remove the watermark.

\noindent $\bullet$ 
 \textit{\textbf{Attack 4 Re-watermarking Attack}}: The adversary dispatches the watermarked texts into another LLM watermarking framework like \sys{} and re-watermark it to remove the inserted signatures.

%Due to the limited available datasets, we do not consider the human rephrase attack, where they can also be added into the training pipeline of Section~\ref{subsec:archi} to make \sys{} more resilient toward such attack.

\section{\sys{} Design} 
\label{sec:method}
\begin{figure*}[!ht]
    \centering
    \includegraphics[width=0.9 \textwidth]{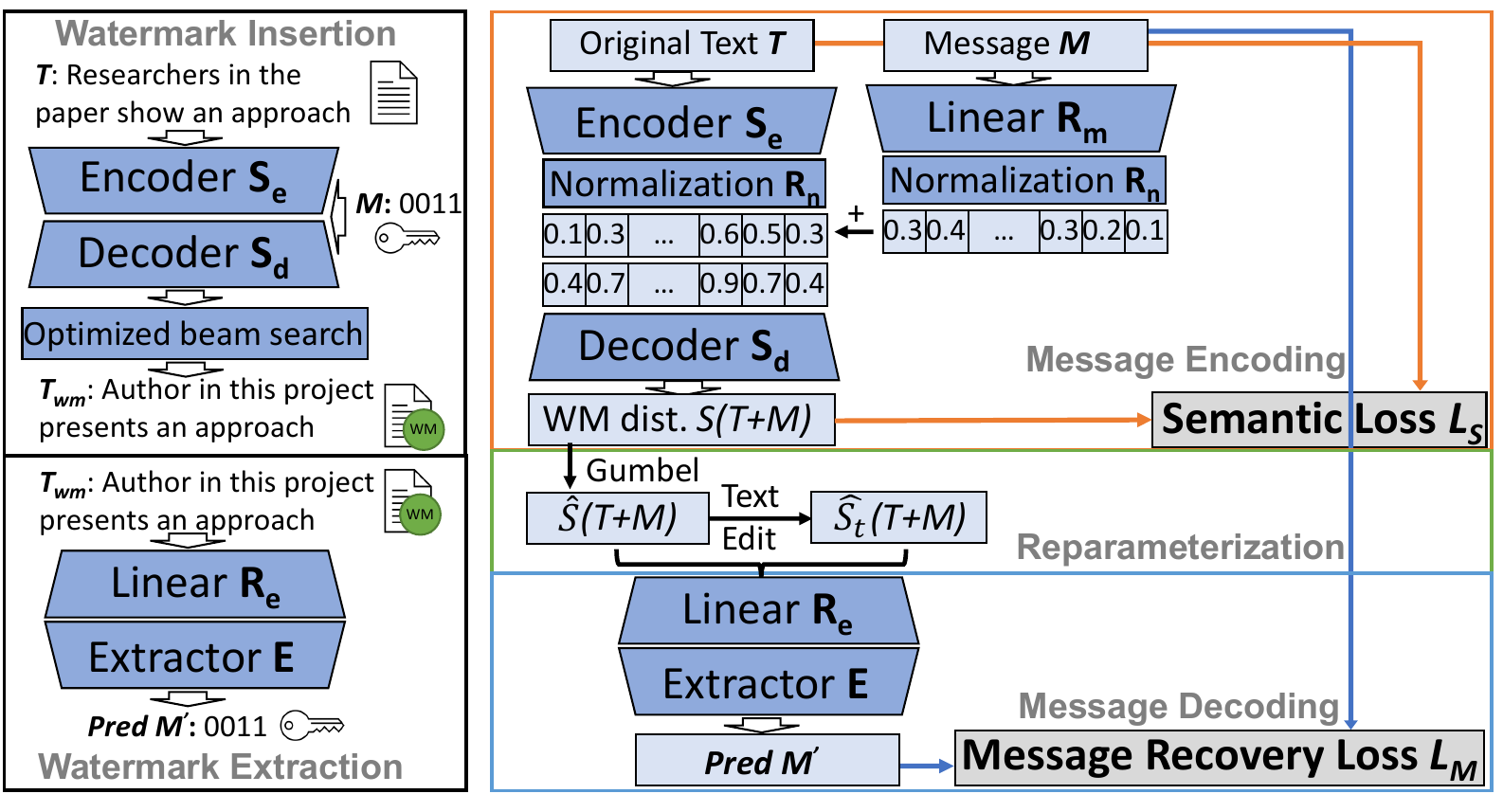}
    \vspace{-10pt}
    \caption{\sys's Watermarking Framework. The left is an overview of \sys{}: The message encoding module leverages an optimized beam search algorithm to produce coherent watermarked contents. The message decoding module is designed for efficient watermark extraction.  The right is \sys's training pipeline: The message encoding, reparametrization, and message decoding modules are trained jointly in an end-to-end fashion, aiming to minimize the semantic loss between original text $T$ and watermarked distribution $S(T+M)$, as well as minimize the message recovery loss between the inserted message $M$ and predicted message $M^\prime$.
    }
    \label{fig:training}
    \vspace{-10pt}
\end{figure*}

In this section, we first introduce \sys's training pipeline as illustrated in Figure~\ref{fig:training}.  Then, with the trained \sys, we introduce how coherent watermarked texts are generated from the message encoding module and how message signatures are extracted from the watermarked texts in the message decoding module.

\subsection{\sys{} Training}\label{subsec:archi}

The watermarking framework is depicted in Figure~\ref{fig:training}, which is a confluence of three major components: \textbf{Message Encoding}, \textbf{Reparameterization}, and \textbf{Message Decoding}. The message encoding inserts invisible watermarks into the LLM-generated texts via a Sequence-to-sequence (Seq2Seq) model.  Reparameterization converts the watermarked distribution from the Seq2Seq model towards a more sparse distribution of the watermarked textual tokens using Gumbel-Softmax.  Message decoding first maps the reparameterized distribution into their respective embedding representation using a mapper network. It then employs a transformer-based decoder to extract the secret messages from the embedding.   

\noindent\textbf{\textit{Message Encoding:}}
The message encoding module takes an LLM-generated token sequence, denoted as $T = \{T_1, T_2, ... T_t\}$, alongside a binary signature sequence $M$ as input.  The texts $T$ goes through the encoder $\mathbf{S}_e$ of the Seq2Seq model and acquires its corresponding latent space representation $\mathcal{S}_e(T)$ at the final normalization layer $\mathbf{R}_n$. Concurrently, the message $M$ is encoded by a linear layer $\mathbf{R}_m$ followed by the shared normalization layer $\mathbf{R}_n$ into the same latent space representation as $R_n(M)$.
At the latent space, the messages are embedded into every token in $T$ as $\mathcal{S}_e(T+M)$. The embedded latent feature $\mathcal{S}_e(T+M)$ is directed to the Seq2Seq's decoder $\mathbf{S}_d$  to obtain the watermarked distribution over the vocabulary as $\mathcal{S}(T+M)$. 

%\begin{equation}\label{eq:addition}
%   \mathcal{S}_{i}(T+M) = S_d (S_e(T_i)+R_n(M)) \quad \text{for } i=1, \ldots, |T|
%\end{equation}

\noindent\textbf{\textit{Reparameterization:}}
The message encoding module generates a dense token distribution, whereas the message decoding module extracts messages from the watermarked textual tokens' one-hot encoding. To bridge this gap, the reparameterization module connects them and transforms the dense token distribution into a sparser form while ensuring differentiability. 
To achieve this transformation, Gumbel-Softmax is applied in Equation~\ref{eq:gumbel} to approximate the watermarked distribution $\mathcal{S}(T+M)$ to more sparse encoding, denoted as $\hat{\mathcal{S}}(T+M)$. $\mathcal{S}(T+M)$ is simplified as $\mathcal{S}$. The $g_i$ is the noise i.i.d samples drawn from Gumbel(0,1), $|V|$ is the vocabulary size, and $\tau$ is the temperature for sampling. The lower $\tau$ is, the closer the reparameterized $\hat{\mathcal{S}}_i$ is to one-hot encoding. 

\begin{equation}
\label{eq:gumbel}
\begin{array}{ll}
\hat{\mathcal{S}}_i=\frac{\exp \left(\left(\log \left(\mathcal{S}_i\right)+g_i\right) / \tau\right)}{\sum_{j=1}^{|V|} \exp \left(\left(\log \left(\mathcal{S}_j\right)+g_j\right) / \tau\right)} 
\quad   \text{for } i=1, \ldots, |T|
 \end{array}
\end{equation}

\noindent\textbf{\textit{Message Decoding:}} 
To decode embedded $M$ from the reparameterized distribution  $\hat{\mathcal{S}}(T+M)$, \sys{} first maps it into the embedding space using a linear layer $\mathbf{R}_e$, yielding $\mathcal{H}(T+M)$. The $\mathcal{H}(T+M)$ is the watermarked text representation in the embedding space. Then, the transformer-based extractor model $\mathbf{E}$ extracts messages from $\mathcal{H}(T+M)$ as $M^\prime = \mathcal{E}({\mathcal{H}(T+M)})$.

\sys{} becomes robust by enforcing the extractor to learn the embeddings of the malicious transformations and decode the same messages $M$ from those transformations as well. 
The transforms, including randomly dropping, adding, and replacing tokens in the watermarked distribution, are performed over $\hat{\mathcal{S}}(T+M)$ and get their corresponding distribution as $\hat{\mathcal{S}}_t(T+M)$. Similar to $\hat{\mathcal{S}}(T+M)$, the $\hat{\mathcal{S}}_t(T+M)$ is mapped to the embedding space and extracts messages as $M_t^\prime =\mathcal{E}({\mathcal{H}_t(T+M)})$.

\noindent\textbf{\textit{End-to-end Training:}}
The above modules are trained in an end-to-end manner, with objectives to (1) ensure the semantic similarity of the input text $T$ and the watermarked distribution $S(T+M)$  and (2) ensure the watermark extraction of the input message $M$ and decoded message $M^\prime$ and $M_t^\prime$.  The first objective is reflected by the semantic loss $L_s$ and the second is reflected by the message recovery loss $L_M$.

\textbf{(1) Semantic Loss}: \sys{} formulates the semantic loss $L_S$  by minimizing the cross entropy loss between input token $T$ and watermarked text distribution $\mathcal{S}(T+M)$  in Equation~\ref{eq:loss_encoding}. To avoid overfitting, in every epoch, the input token sequence $T$ is randomly masked via a mask sequence $T_M$. $T_M$ is of the same size as $T$, where 1 means the token is unmasked and 0 means the token is masked.
$|V|$ is the size of vocabulary in $\mathcal{S}$ and $|T|$ is the number of tokens in the input text $T$.

\begin{equation}
\label{eq:loss_encoding}
L_S(T, \mathcal{S}(T\cdot T_M+M)) = -\frac{1}{|T|}  \sum_{i=1}^{|T|}\sum_{j=1}^{|V|} T_{ij} \log(\mathcal{S}_{ij}(T\cdot T_{M}+M)) 
\end{equation}

\textbf{(2) Message Recovery Loss}:  \sys{} measures the message recovery loss $L_M$ between the input message  $M$ and decoded message $M^\prime$ from the watermarked distribution using $L_1$ loss. Similarly, \sys{} also includes the message recovery loss between the signatures decoded from malicious transformation $M_t^\prime$ and input message  $M$. In Equation~\ref{eq:loss_decoding}, the two losses are adjusted by the coefficients $w_w$ and $w_t$.

\begin{equation}
\label{eq:loss_decoding}
\begin{array}{ll}
L_M(M, M^\prime,  M_t^\prime) = w_w \sum_{i=1}^{|M|} |M_i - M^\prime| + w_t \sum_{i=1}^{|M|} |M_i - M_t^\prime|
\end{array}
\end{equation}

\textbf{(3) Training}:  When training the end-to-end framework, we include the above losses together as an objective function in Equation~\ref{eq:total}. 
This is reflected in \sys{}, where (i) minimizing $\mathcal{L}_S$ means encouraging the watermarked texts to be semantically close to the input texts, and (ii) minimizing $\mathcal{L}_M$ to ensure the encoded messages can be successfully extracted from the watermarked texts. The $w_1$ and $w_2$ are the trade-off coefficients during training.  

\begin{equation}
\label{eq:total}
L = w_1  L_S + w_2 L_M
\end{equation}

\subsection{Watermark Insertion}
With a trained \sys, model proprietors can use our message encoding module to insert watermarks into the LLM-generated contents. The message encoding takes the LLM-generated text $\bar{T}$ and the message $\bar{M}$ as input and generates the watermarked distribution over the vocabulary as $\mathcal{S}(\bar{T}+\bar{M})$. 
Decoding the distribution $\mathcal{S}(\bar{T}+\bar{M})$ by simply taking its maximum index without considering the overall sentence structure will diminish the text coherence. To overcome this, an optimized beam search algorithm is introduced in Algorithm~\ref{alg:insertion}. It aims to ensure coherence while maximizing the watermark extraction rates.

In every decoding step, a Gumbel-Softmax noise with temperature $\tau$ is added into the token distribution $\mathcal{S}_i$. Then, the beam search algorithm with beam size $B$ produces $B$ candidate sentences from the perturbed token distribution. For each sentence, \sys{} evaluates their extraction accuracy from the extractor in the message decoding module. \revision{Based on the empirical evidence, we find a small beam size $B$ results in readable watermarked texts, whereas the best-accuracy sentence has good watermark extractability} The beam search is repeated for $K$ iterations with different temperatures $\tau_k$ to obtain more diverse watermarked texts.
 
\begin{algorithm}[!ht]
\caption{Optimized Beam Search Algorithm}
\label{alg:insertion}
\begin{algorithmic}[1]
\Require LLM-generated text token $\bar{T}$, temperature list $\tau$, beam size $B$, number of iterations $K$, message $\bar{M}$
\Ensure Watermarked text $\bar{T}_{wm}$ 
\State Initialize \text{max\_accuracy} = 0
\State Initialize $\bar{T}_{wm}$  = None
\For {$k$ = 1 to  $K$}
\State  Initialize mask $\bar{T}_M$
\State  Initialize watermarked dist. $\mathcal{S}(\bar{T}\cdot\bar{T}_M+\bar{M})$
\For {each $\mathcal{S}_i$  in  $\mathcal{S}(\bar{T}\cdot\bar{T}_M+\bar{M})$}
        \State  $\mathcal{S}_{\text{noisy},i} \leftarrow \mathcal{S}_{i} + \text{Gumbel}(\mathcal{S}_{i}, \tau_k) $
\EndFor
\State $T_k \leftarrow \text{Beam\_Search}(\mathcal{S}_{\text{noisy}}, B)$ 
\For {each  $T_{ki}$ in $T_k$}
\State  $a \leftarrow \text{Accuracy}(\mathbf{E}(T_{ki}), \bar{M})$
\If { $a > \text{max\_accuracy} $}
    \State $\text{max\_accuracy} \leftarrow a$
    \State $\bar{T}_{wm} \leftarrow T_{ki}$
\EndIf
\EndFor
\EndFor \\
\Return $\bar{T}_{wm}$ 
\end{algorithmic}
\end{algorithm}
 
%To get a more diverse decoded sentence, we add Gumbel noise with temperature $\bar{\tau}$ into every token of the $\mathcal{M} = N \times |\bar{T}|$ matrix and get the noisy $\bar{\mathcal{M}}$. Then, we apply beam search with beam size $B$ to extract top $B$ probability sentences from the distribution $\bar{\mathcal{M}}$.  The process is repeated $k$ times and gets $k\times B$ sentences.  
%We apply the fine-tuned extractor $\mathbf{E}$ from to extract messages from the $k\times B$ sentences and return the one with the highest extraction rates as watermarked text $\bar{T}_{wm}$.

%A small beam size $B$ ensures the resultant texts are highly readable, whereas the selected best-accuracy sentence guarantees the watermark extractability. T

\subsection{Watermark Extraction}\label{subsec:wm_extract}

\sys{} extracts the watermark via the message decoding module. Given the watermarked text $\bar{T}_{wm}$, it is first mapped into the embedding space using $\mathbf{R}_e$. Then, $\mathbf{E}$ extracts the predicted message $\bar{M}^\prime$ from $\bar{T}_{wm}$ and compares it with LLM proprietor inserted watermark $\bar{M}$ to claim ownership.

\noindent\textbf{Watermark Strength} The confidence in predicting if watermark signatures reside in the watermarked texts can be evaluated using the z-score. The larger the z-score is, the more robust protection the watermark can provide. 
Given a message sequence with length $|M|$, $|N|$ bits out of the message can be successfully detected. The message generation is random and follows binomial distributions \revision{as in \baseend~\cite{abdelnabi21oakland}}, where the probability for generating bit 0 is $p=0.5$ and bit 1 is $1-p = 0.5$.  The mean of the message distribution can be calculated as $\mu= |M| \times p$, and the variance can be calculated as $ \sigma^2= |M| \times p \times(1-p)$. We calculate the z-score of the binominal distribution in Equation~\ref{eq:zscore}.

\vspace{-5pt}
\begin{equation}
\begin{aligned}\label{eq:zscore}
z=\frac{|N|-\mu}{\sigma}
\end{aligned}
\end{equation}
 \vspace{-5pt}

%p-value. In Null Hypothesis, a large p-value (e.g. larger than 0.05) means the observed data is consistent with the null hypothesis, meaning the watermarked texts and non-watermarked texts cannot be significantly distinguished. A small p-value means the the Null Hypothesis is rejected and the watermark insertion confidence is high. 

%Inserting watermark signatures into LLM-generated texts is independent of the inference stage.  
%The p-value of the hypothesis test is calculated as Equation~\ref{eq:pvalue}, where the extracted number of bits higher than $k$ is considered as the watermark is successfully extracted. $n$ is the length of watermark signatures.

%\begin{equation}\label{eq:pvalue}
%    P_c = \sum_{i=k}^n\left(\begin{array}{l}
%n \\
%i
%\end{array}\right) 0.5^n
%\end{equation}

%If $m$ text segments are stacked together and form the paragraph sequence as $T_s = \{T_{s1}, T_{s2}, ...T_{sm}\}$. The watermark can be further strengthened by computing as $P_c = P_{c1} \cdot P_{c2} \ldots P_{cm}$.

%We calculate the z-score, if $n$ 

\section{Experiments}
\label{sec:exp}

In this section, we first introduce the experiment setup. Then, we demonstrate \sys{}'s effectiveness, efficiency, and transferability compared with prior arts. Next, we present an ablation study on the effectiveness of each component in \sys. Finally, we analyze \sys's robustness by evaluating its performance under a spectrum of watermark removal and detection attacks.  
\subsection{Experiments Setup}

\noindent\textbf{Datasets}
We use four datasets to benchmark the LLM-generated content watermarking performance. The HC3~\cite{guo-etal-2023-hc3} is the ChatGPT-generated response to questions from QA platforms (e.g., Quora and Stack Overflow). The Human Abstract~\cite{sivesind_2023} and ChatGPT Abstract~\cite{sivesind_2023}  are the research abstracts written by human researchers and their rephrased version by GPT-3.5 Turbo.  The WikiText-2~\cite{merity2016pointer}  is a collection of paragraphs extracted from verified Good and Featured articles on Wikipedia. The detailed statistics are summarized in Table~\ref{tab:coeff}. We randomly split 80\% of the texts as training datasets, and the remaining 20\% are test datasets for HC3, ChatGPT Abstract, and Human Abstract.

%and ChatGPT Abstract~\cite{sivesind_2023} are LLM-generated, and the Human Abstract~\cite{sivesind_2023} and WikiText-2~\cite{merity2016pointer} are human-written.  

\begin{table}[h]
\centering
\vspace{-5pt}
\resizebox{\columnwidth}{!}{
\begin{tabular}{cccc}
\toprule
Dataset & Train & Test & Data Source \\ \midrule
HC3~\cite{guo-etal-2023-hc3} & 19440 & 4860 & LLM \\  
WikiText-2~\cite{merity2016pointer} & 44800 & 4360 & Human \\
ChatGPT Abstract~\cite{sivesind_2023} & 8000  & 2000 & LLM \\  
Human  Abstract~\cite{sivesind_2023} & 8000 & 2000 & Human \\  
\bottomrule
\end{tabular}}
\caption{Dataset to benchmark the watermarking performance.\label{tab:coeff}}
 \vspace{-10pt}
\end{table}

\begin{table*}[!ht]
  \centering
 \resizebox{\textwidth}{!}{%
  \begin{tabular}{c|c|ccc|ccc|ccc}
    \toprule
\multirow{2}{*}{Dataset} & \multirow{2}{*}{Methods}  &\multicolumn{3}{c|}{4 bits} & \multicolumn{3}{c|}{8 bits} & \multicolumn{3}{c}{16 bits}\\
  &    &  WER(\%) $\uparrow$ & BERT-S $\uparrow$ &BLEU-4 $\uparrow$ &  WER(\%) $\uparrow$ & BERT-S $\uparrow$ &BLEU-4 $\uparrow$&  WER(\%) $\uparrow$ & BERT-S $\uparrow$ &BLEU-4 $\uparrow$\\
    \midrule
 \multirow{3}{*}{HC3}
&  \sys & \underline{97.01} & \underline{0.93} & \underline{0.43}& \underline{95.59}  & \underline{0.92}&\underline{0.45} & \underline{73.46}&\textbf{0.92} & \textbf{0.46}\\
&  \baseend~\cite{abdelnabi21oakland} & 96.32&\textbf{0.94} & \textbf{0.91} &74.08 & \textbf{0.96}& \textbf{0.84} & \textcolor{red}{50.00} & - & - \\
&   \baseinfer~\cite{kirchenbauer2023watermark} & \textbf{99.43} & 0.62 &  \textcolor{red}{0.01} & \textbf{99.43} & 0.62 &  \textcolor{red}{0.01}  & \textbf{99.43} & 0.62 &  \textcolor{red}{0.01}  \\
   \midrule
  \multirow{3}{*}{WikiText-2}
&   \sys & \textbf{97.23} & \underline{0.92} &  \underline{0.33} & \underline{89.57}& \underline{0.89} & \underline{0.23}& \underline{76.37} & \textbf{0.89} & \textbf{0.19} \\
&  \baseend~\cite{abdelnabi21oakland} & 97.04 & \textbf{0.95}& \textbf{0.92} & 78.18 & \textbf{0.96} & \textbf{0.92}&  \textcolor{red}{50.00}  & - & - \\
&  \baseinfer~\cite{kirchenbauer2023watermark} & \underline{97.07}  & 0.65 & \textcolor{red}{0.01} & \textbf{97.07}  & 0.65 & \textcolor{red}{0.01} & \textbf{ 97.07}  & 0.65 & \textcolor{red}{0.01} \\
%&   \baserule~\cite{he2022cater} & \\
  \midrule
  \multirow{3}{*}{\begin{tabular}[c]{@{}c@{}}ChatGPT\\Abstract\end{tabular} }
&   \sys & \underline{96.98} &  \underline{0.91} & \underline{0.30}  & \underline{93.53} & \underline{0.91} & \underline{0.29} & \underline{73.80}  & \textbf{0.90} & \textbf{0.24}\\
&  \baseend~\cite{abdelnabi21oakland} & 83.81 & \textbf{0.94} & \textbf{0.96} & 62.28 & \textbf{0.96} & \textbf{0.83} & \textcolor{red}{50.00}   & - & - \\
&  \baseinfer~\cite{kirchenbauer2023watermark} & \textbf{99.87} & 0.63 & \textcolor{red}{0.01}& \textbf{99.87} & 0.63 & \textcolor{red}{0.01} & \textbf{99.87} & 0.63 & \textcolor{red}{0.01}\\      
%&   \baserule~\cite{he2022cater} & \\
  \midrule
  \multirow{3}{*}{\begin{tabular}[c]{@{}c@{}}Human\\Abstract\end{tabular} }
&   \sys &  \underline{96.85} & \underline{0.89} & \underline{0.26}  & \underline{88.85}& \underline{0.88}&\underline{0.20} & \underline{75.81} & \textbf{0.84} & \textbf{0.10}\\
&  \baseend~\cite{abdelnabi21oakland} & 71.39 & \textbf{0.95} &  \textbf{0.95}& 63.78& \textbf{0.95} & \textbf{0.85} & \textcolor{red}{50.00} & - & - \\
&  \baseinfer~\cite{kirchenbauer2023watermark} & \textbf{99.50} & 0.68 & \textcolor{red}{0.01}& \textbf{99.50} & 0.68 & \textcolor{red}{0.01} & \textbf{99.50} & 0.68 & \textcolor{red}{0.01}\\    
%&   \baserule~\cite{he2022cater} & \\
    \bottomrule  
  \end{tabular}
  }
\vspace{-10pt}
  \caption{Segment-level watermarking comparison. The length of the segment is 80 tokens. Both \sys{} and \baseend{} are trained on HC3 and WikiText-2's training dataset and report the watermarking performance on the test dataset. The transferability is benchmarked by reporting the test accuracy on ChatGPT Abstract with HC3-trained frameworks and on Human Abstract with WikiText-2-trained frameworks. The best metric values are highlighted in \textbf{bold} text, and the second best metric values are \underline{underlined}. Metric values that are highlighted in \textcolor{red}{red} suggest failure cases (low WER or high semantic distortion). A WER of 50\% indicates watermark recovery failure. The WER for all unwatermarked texts generated by the original LLM stands at 50\%.
  \vspace{-10pt}
  % signifies that the framework has failed to successfully embed watermarks of the specified bit length.
  % WER for all unaltered texts generated by the original LLM stands at 50\%, which is not shown in the table. 
  % and we skip them in the table.  
  \label{tab:effective}}
 
\end{table*}
\noindent\textbf{Evaluation Metrics}
\revision{We benchmark \textbf{ 
Effectiveness} by the fraction of the inserted watermarks successfully extracted, reflected by the Watermark Extraction Rate. \textbf{Fidelity} is measured by (1) BERT-S~\cite{zhang2019bertscore}, indicating the watermarked texts should be semantically similar to the original texts, and (2) BLEU-4~\cite{papineni2002bleu} that the watermarked texts should be coherent and consistent w.r.t. the original texts. We compute these metrics using Huggingface Evaluate~\cite{wolf-etal-2020-transformers} }and the measurement details are as follows:

%We benchmark the effectiveness of the watermarking framework from three aspects: (1) The inserted watermarks should be successfully extracted as reflected by the Watermark Extraction Rate; (2) The watermarked texts should be semantically similar to the original texts as reflected by the BERT-S~\cite{zhang2019bertscore}; and (3) the watermarked texts should be coherent and consistent w.r.t. the original texts as reflected by BLEU-4~\cite{papineni2002bleu}. %and (4) the watermarked texts should be fluent as reflected by Perplexity~\cite{popel2010perplexity}~\footnote{https://huggingface.co/docs/transformers/perplexity} at the document level.   %It measures the lexical similarity via n-grams match

\noindent $\bullet$ \textit{\textbf{Watermark Extraction Rate (WER)}}: the percentage of the binary watermark message successfully extracted. 

\noindent $\bullet$ \textit{\textbf{BERT-S}}~\cite{zhang2019bertscore}: the BERT score cosine distance between the original and watermarked text.

\noindent $\bullet$ \textit{\textbf{BLEU-4}}~\cite{papineni2002bleu}: the number of n-grams(n=4) in the watermarked text that match the reference texts.
 %  \item Perplexity (PPL)~\cite{popel2010perplexity}: the exponentiated average negative log-likelihood of the watermarked text.  

\revision{BLEU-4 is adopted from machine translation to measure the n-gram(n=4) matches between the translated and baseline texts. While the metric has no hard threshold, baselines in prior work~\cite{wu2019machine} indicate BLEU-4 of higher than 0.15 is considered a semantically coherent document transformation.}

\textbf{Efficiency} of the watermarking frameworks are measured from two aspects: (1) the time overhead for inserting watermarks and (2) the computation resources required for inserting watermarks, as reflected by the peak GPU memory. Evaluations for \textbf{Robustness} and \textbf{Undetectability} are in Section~\ref{subsec:attack}.

\noindent\textbf{Baselines}
We compare \sys{} with four state-of-the-art LLM watermarking frameworks: \baserule~\cite{he2022cater}, \baseinfer~\cite{kirchenbauer2023watermark}, \baseinferrb~\cite{kuditipudi2023robust} and \baseend~\cite{abdelnabi21oakland}. 
The rule-based watermarking algorithm \baserule~\cite{he2022cater} inserts conditional watermarks into the LLM-generated texts. It inserts watermarks by choosing words that minimize the distortion of overall word distributions while maximizing the change of conditional word selections. 
The inference-time watermarking approach \baseinfer~\cite{kirchenbauer2023watermark} inserts watermarks at the LLM decoding step. By dynamically splitting the vocabulary into green/red lists with the message signatures, it enforces the next token prediction to only sample from the green list. \revision{We employ the soft red list watermarking algorithm in KGW~\cite{kirchenbauer2023watermark}. The hyperparameters and the prompt methodologies follow the default settings.} \baseinferrb~\cite{kuditipudi2023robust} tries to reduce the semantic distortion by proposing an exponential minimum sampling strategy at the LLM decoding stage. 
The neural-based watermarking approach \baseend{} watermarks LLM-generated texts in an end-to-end manner. It trains a transformer-based encoder-decoder network that takes an input sentence and a binary message to produce a watermarked text.  \baseend{} is trained to preserve the semantics of the watermarked texts while ensuring signature extraction.

For fair comparisons, we compare \sys{} with baselines at the text segment level with a fixed 80 token length \revision{following \baseinferrb~\cite{kuditipudi2023robust} and \baseend~\cite{abdelnabi21oakland}. We use the long text sequences with a fixed 640 token length as a proof-of-concept showing \sys{} can watermark longer sequences, which exceeds the maximum length studied in prior work. }
\baseinfer{}~\cite{kirchenbauer2023watermark} and \baseinferrb{}~\cite{kuditipudi2023robust} use OPT-2.7B~\cite{zhang2022opt} as the backbone generator to insert watermarks. \revision{For \baserule~\cite{he2022cater}, we employ their open-sourced synonym tables in follow-up work~\cite{he2021protecting}.} For \baseinfer{}, we follow their paper setting~\cite{kirchenbauer2023reliability} and consider the watermarking to be successful if the z-score of the watermark extraction exceeds 4. For \baseinferrb{}, the z-scores are below 4 after watermarking, and we report the WER providing the same level of p-value strength as \baseinferrb{} in Table~\ref{tab:effective-paragraph}.
%As \baserule~\cite{he2022cater} did not open-source their synonym tables and optimization functions for transformation upon our submission, we directly use the numbers they reported in the original paper and add them in Table~\ref{tab:effective-paragraph}. 

% At the segment level, to insert $|M|$ message bits into \baseinfer{}, we randomly select $|M|$ indexes from the 80 tokens and split the green/red list only in those $|M|$ indexes. 
%\textcolor{red}{TO SH: 1. Do you think we should compare with rule-based watermarking approaches?}

\noindent\textbf{Hyperparameter Settings} We include more information on \sys's training hyperparameters and architecture details in Appendix~\ref{append_setup}.

\subsection{Results}
In this subsection,  we demonstrate \sys's effectiveness and efficiency. The robustness evaluations are in Section~\ref{subsec:attack}.

\subsubsection{Segment-level Watermarking}

We summarize the segment-level watermarking performance in Table~\ref{tab:effective}, where the unit segment length is 80 tokens following \baseend~\cite{abdelnabi21oakland}. 
For LLM-generated texts, \sys{} and \baseend{} are trained on the HC3 training set and report the performance on the test set of HC3 and ChatGPT Abstract. For human-written texts, \sys{} and \baseend{} are trained on the WikiText-2 training set and report the performance on the test set of WikiText-2 and Human Abstract. For \baseinfer{}, we use the first 40 tokens as prompts, and the predicted next 80 tokens are as watermarked texts. The inserted signature length is increased from 4-bit to 16-bit. All the original LLM-generated texts' WER are 50\%, and we skip them in the table. 

\noindent\textbf{Comparison with \baseend}~\cite{abdelnabi21oakland}:
(1) \sys{} can insert more signature bits into the same text length without compromising the semantics. The \baseend's WER dropped to $\sim$ 70\% when inserting 8 bits signatures into the token sequence and failed to insert watermarks larger than 16 bits.  However, \sys{} can extract more than 90\% of the signature when 8 bits are inserted. (2)  \baseend{} demonstrates worse transferability compared with \sys{}. When 4 bits signatures are inserted into 80 tokens, the WER of \baseend{} trained on WikiText-2 drops 25.65\% when inference on the Human Abstract dataset and drops 12.51\% from HC3 to ChatGPT Abstract dataset.  For \sys{}, the WER only drops 0.38\% and 0.03\% from WikiText-2 to Human Abstract and from HC3 to ChatGPT Abstract, respectively. (3) \baseend{} achieves higher BLEU-4 by replacing words with their synonyms but does not modify the syntactic structure of the sentences. This hinders \baseend{} from being robust under more powerful rephrase attacks, as shown in Section~\ref{subsec:attack}. By applying masking strategies during \sys{} training, it achieves similar semantic preservation as \baseend{}, but more diverse outputs as reflected by lower BLEU-4. 

\noindent\textbf{Comparison with \baseinfer}~\cite{kirchenbauer2023watermark}: (1) \baseinfer{} demonstrates better WER compared with \sys{} by inserting watermarking at the LLM decoding step. However, this is at the cost of compromising its semantics and coherence. (2) \baseinfer's average semantics is dropped by 29\% compared with \sys{} and  \baseinfer's coherence score BLEU-4 is close to 0. This demonstrates the \baseinfer{} significantly distorts the semantics of the original LLM-generated texts and adversely compromises the LLM-generated content quality.

%\baseinfer{} can successfully insert and extract watermarks into LLM-generated texts with WER up to 99\%. However, watermarked texts and original LLM-generated texts' semantics become dramatically different as more signature bits are inserted.  While the maximum number of bits \sys{} can insert is 8 bits, the watermark insertion process does not compromise the watermarked text semantics. 

\subsubsection{Watermarking Long Sequences}
The responses local users receive from the LLMs are generally long text sequences with multiple paragraphs. In this section, we investigate how effectively long sequences can be watermarked with the aforementioned frameworks. 
For the long sequence watermarking, we set the maximum token length to 640 as a proof-of-concept showing \sys{} can watermark longer sequences, which exceeds the maximum length studied in prior work.  As \baseend{} performs watermarking at the segment level, we report the watermarking performance by dividing the long sequence into multiple segments and watermarking each segment individually. We stop watermarking later segments if the maximum length is smaller than 640 tokens. 
We train both \baseend{} and \sys{} on the HC3 dataset and report the inference watermarking performance on all four datasets with the trained framework. The results are reported in Table~\ref{tab:effective-paragraph}. The WER from unwatermarked content generated by LLMs is 50\%. Therefore, we skip it in the table.

% All the original LLM-generated texts' WER are 50\%. 

%The \sys-paragraph refers to the paragraph watermarking, and the \sys-segment refers to the stacked segment watermarking like \baseend. 

\begin{table*}[!ht]
  \centering
 \resizebox{\textwidth}{!}{%
  \begin{tabular}{c|c|ccc|ccc|ccc}
    \toprule
\multirow{2}{*}{Dataset} & \multirow{2}{*}{Methods}  &  \multicolumn{3}{c|}{16 bits} & \multicolumn{3}{c|}{32 bits}& \multicolumn{3}{c}{64 bits}\\
   &    &  WER(\%) $\uparrow$  & BERT-S $\uparrow$ &BLEU-4 $\uparrow$ & WER(\%) $\uparrow$  & BERT-S $\uparrow$ &BLEU-4 $\uparrow$ & WER(\%) $\uparrow$  & BERT-S $\uparrow$ &BLEU-4 $\uparrow$ \\
    \midrule
 \multirow{5}{*}{HC3}
%&  \sys-P  & \\
&  \sys  &\underline{98.93} & \underline{0.94}& \underline{0.45} & \underline{97.84} & \underline{0.92} & \underline{0.41} & \underline{95.61} & \underline{0.91} & \underline{0.41} \\
&  \baseend~\cite{abdelnabi21oakland} & 96.12& \textbf{0.98} & \textbf{0.95}& 94.51 & \textbf{0.98} & \textbf{0.93} & 68.42 & \textbf{0.95} & \textbf{0.82} \\
&   \baseinfer~\cite{kirchenbauer2023watermark} & \textbf{99.57} & 0.58 & \textcolor{red}{0.01} &  \textbf{99.57} & 0.58 & \textcolor{red}{0.01}  & \textbf{99.57} & 0.58 & \textcolor{red}{0.01} \\
& \baseinferrb~\cite{kuditipudi2023robust} & 79.37& 0.80 & 0.01 & 70.62 & 0.80 & 0.01 & 64.68 & 0.80 & 0.01 \\
%&   \baserule~\cite{he2022cater} & 100.00 & 0.65 & 0.30 & 93.75 & 0.65 & 0.30 & 81.25 & 0.65 & 0.30  \\
&   \baserule~\cite{he2022cater} & \revision{75.20} & \revision{0.96} & \revision{0.63} & \revision{75.20} & \revision{0.96} & \revision{0.63} & \revision{75.20} & \revision{0.96} & \revision{0.63} \\
\midrule
  \multirow{5}{*}{WikiText-2}
%&  \sys-P  & \\
&  \sys  & \underline{99.02}& \underline{0.92} & \underline{0.32} & \underline{98.60}& \underline{0.86} & \underline{0.18} &  \underline{94.48} & \underline{0.85} & \underline{0.16} \\
&  \baseend~\cite{abdelnabi21oakland} & 89.72& \textbf{0.97} & \textbf{0.95} &  85.82& \textbf{0.95} &\textbf{0.93} & 65.77 & \textbf{0.96} &  \textbf{0.85} \\
&  \baseinfer~\cite{kirchenbauer2023watermark} & \textbf{99.13} & 0.61& \textcolor{red}{0.02}  &\textbf{99.13} & 0.61&\textcolor{red}{0.02}   &\textbf{99.13} & 0.61& \textcolor{red}{0.02}    \\
& \baseinferrb~\cite{kuditipudi2023robust} & 79.37& 0.82 & 0.01 & 70.62 & 0.82 & 0.01 & 64.68 & 0.82 & 0.01 \\
&   \baserule~\cite{he2022cater} & \revision{53.10} & \revision{0.94} & \revision{0.75} & \revision{53.10} & \revision{0.94} & \revision{0.75} & \revision{53.10} & \revision{0.94} & \revision{0.75}\\
%&   \baserule~\cite{he2022cater} & 100.00 & 0.65 & 0.30 & 93.75 & 0.65 & 0.30 & 81.25 & 0.65 & 0.30\\53.10%/0.94/0.75
  \midrule
  \multirow{5}{*}{\begin{tabular}[c]{@{}c@{}}ChatGPT\\Abstract\end{tabular} }
%&  \sys-P  & \\
&  \sys  & \underline{98.24} & \underline{0.92}& \underline{0.33}& \underline{98.55} & \underline{0.90}  & \underline{0.27}& \underline{95.04} & \underline{0.89}   & \underline{0.27}  \\
&  \baseend~\cite{abdelnabi21oakland} & 88.26& \textbf{0.96} & \textbf{0.95}&  80.62 & \textbf{0.97} & \textbf{0.94}  & 62.39 & \textbf{0.95}& \textbf{0.84}  \\
&  \baseinfer~\cite{kirchenbauer2023watermark} &\textbf{99.01} &  0.61 &\textcolor{red}{0.01} &\textbf{99.01} &  0.61 &\textcolor{red}{0.01} &\textbf{99.01} &  0.61 &\textcolor{red}{0.01} \\
& \baseinferrb~\cite{kuditipudi2023robust} &79.37 & 0.80 & 0.01 &  70.62& 0.80 & 0.01 & 64.68  & 0.80 & 0.01   \\
&   \baserule~\cite{he2022cater} & \revision{75.50} & \revision{0.93} & \revision{0.64} & \revision{75.50} & \revision{0.93} & \revision{0.64} & \revision{75.50} & \revision{0.93} & \revision{0.64}\\
%& 100.00 & 0.65 & 0.30 & 93.75 & 0.65 & 0.30 & 81.25 & 0.65 & 0.30 \\75.50%/0.93/0.64
  \midrule
  \multirow{5}{*}{\begin{tabular}[c]{@{}c@{}}Human\\Abstract\end{tabular} }
%&  \sys-P  & \\
&  \sys  & \underline{98.56} & \underline{0.91} & \underline{0.31}& \underline{98.71} &  \underline{0.88} & \underline{0.16}  & \underline{95.39} &  \underline{0.87}& \underline{0.15}\\
&  \baseend~\cite{abdelnabi21oakland}  & 86.43& \textbf{0.96} & \textbf{0.93}  & 77.21  &  \textbf{0.98} & \textbf{0.92} & 63.52& \textbf{0.94} & \textbf{0.85}  \\
&  \baseinfer~\cite{kirchenbauer2023watermark} &\textbf{98.79} & 0.69 & \textcolor{red}{0.01} & \textbf{98.79} & 0.69 & \textcolor{red}{0.01} &\textbf{98.79} & 0.69 & \textcolor{red}{0.01}\\
& \baseinferrb~\cite{kuditipudi2023robust} &  79.37&0.81 &0.01& 70.62  &0.81 & 0.01&  64.68  &0.81 & 0.01 \\
&   \baserule~\cite{he2022cater} & \revision{82.00} & \revision{0.95} & \revision{0.54} & \revision{82.00} & \revision{0.95} & \revision{0.54} & \revision{82.00} & \revision{0.95} & \revision{0.54}\\
%&   \baserule~\cite{he2022cater}& 100.00 & 0.65 & 0.30  & 93.75 & 0.65 & 0.30  & 81.25 & 0.65 & 0.30  \\82.00%/0.95/0.54,
    \bottomrule  
  \end{tabular}
  }
\vspace{-10pt}
  \caption{Long text sequence watermarking comparison. The length of the sequence is 640 tokens. The frameworks are trained on HC3's training dataset with the watermarking performance reported on the test dataset. The transferability is benchmarked by reporting the test accuracy on the rest of the datasets with the trained frameworks. The best metric values are highlighted in \textbf{bold} text, and the second best metric values are \underline{underlined}. Metric values that are highlighted in \textcolor{red}{red} suggest failure cases (low WER or high semantic distortion).  The WER for all unwatermarked texts generated by the original LLM stands at 50\%.  \label{tab:effective-paragraph}}
 \vspace{-10pt}
\end{table*}

\noindent\textbf{Comparison with \baseend}~\cite{abdelnabi21oakland}: (1) \baseend{} successfully preserves the semantics and coherence of the LLM-generated contents. However, the WER drops by an average of 19.52\% when the signature length is extended to 64 bits. Thereby, \baseend{}'s ability to embed stronger watermarks into the sequence is significantly hindered. (2) Similar to segment level watermarking, \baseend{} exhibits worse transferability compared with \sys{}. When inserting 32-bit signatures, \baseend's WER dropped by 14\% when watermarking the ChatGPT Abstract dataset using the watermarking model trained on HC3. However, the WER of \sys{} from HC3 to ChatGPT Abstract showcased no accuracy drop.

% While \baseend preserves the semantics and fluency of the overall paragraph, it fails to insert the longer 64-bit sequences and the overall WER is lower than 70\%. For \sys, both at the segment level and the paragraph level, it remains highly extractable and does not compromise the semantics and fluency with 64-bit insertion. 

\noindent\textbf{Comparison with \baseinfer}~\cite{kirchenbauer2023watermark}: (1) The semantic distortion introduced by \baseinfer{} becomes worse for longer text sequence. For the LLM-generated content, \baseinfer's BERT-S drops by 35\% compared with \sys{}. This suggests that \baseinfer{} could not preserve the original content quality generated by LLM, resulting in an ineffective watermark insertion.  (2) Besides, the BLEU-4 of \baseinfer{} is close to 0, which demonstrates \baseinfer{} failed to maintain the coherence between the original LLM-generated and watermarked texts. 
%(2) \baseinfer{} demonstrates 20-30 unit lower perplexity compared with \sys{}, but it does not signify \sys{} is less fluent. Perplexity is an approximate measure of the alignment between an LLM's predictive probability distribution and the actual word distribution in the text. As \baseinfer's watermarked texts directly originated from the LLM, PPL can reliably reflect text fluency. Conversely, the texts from \sys{} are the productions of the Seq2Seq model, and some of the benchmarking datasets are human-written. Perplexity cannot be a faithful measure of the watermarked text in this setting. Following previous AWT~\cite{abdelnabi21oakland} and CATER~\cite{he2022cater}, we use BLEU-4 to assess the text fluency and coherence. 
 
%However, this does not necessarily mean \sys{} is less fluent, given its high BLEU-4 score. Conversely, because \baseinfer{} and \baseinferrb{} generates watermarked texts 

%The semantic difference between the original LLM-generated texts and the watermarked texts becomes larger at the paragraph level. Our results indicate that by enforcing the green/red list splitting, \baseinfer{} fails to reach the requirement for the effectiveness of the watermark insertion. Besides, \sys{} provides sufficient watermarking strength and preserves the semantics as shown in Section~\ref{subsec:strength}.  

\noindent\textbf{Comparison with \baseinferrb}~\cite{kuditipudi2023robust} (1) While \baseinferrb{} maintains the semantics integrity for inference-time watermarking, the BERT-S and BLEU-4 are averagely 0.07 and 0.25 lower compared with 64-bit \sys{}. This indicates neural-based \sys{} preserve better semantics and coherence compared with \baseinferrb{} for longer text sequences. (2) \baseinferrb{} preserves the semantic at the cost of weaker watermark insertion. The average p-value \baseinferrb{} inserted into the watermarked texts is 9.9$\times10^{-3}$ and equals an average z-score of 2.36. However, the average z-score \sys{} can provide at 64-bit signature is 7.12. Therefore, \sys{} demonstrates better semantic preservation and stronger watermark insertion than \baseinferrb.

%\baseinferrb{} achieves a similar level of semantic similarity and coherence compared with neural-based approaches like \baseend{} and \sys{}. This is a trade-off of the maximum watermark strength that can be inserted into LLM-generated texts. In Table~\ref{tab:effective-paragraph}, we follow \baseinfer{} and set a hard threshold of a watermark z-score higher than 4 as a successful extraction. The WER becomes close to 0.  The corresponding p-value of \baseinferrb{} is 9.9$10^{-3}$, which is smaller than 0.25. It can showcase watermarks residing in the texts but cannot reach the level of successful decoding and proof the watermarking signature is exactly inserted by the LLM owner. 

\noindent\textbf{Comparison with \baserule}~\cite{he2022cater}: 
\revision{\baserule{} achieves high semantic preservation by replacing words with their synonyms. However, such replacements are not generalizable toward new datasets because the candidate words can have high frequency on certain datasets but low frequency on the rest. As in Table~\ref{tab:effective-paragraph}, \baserule{} has higher WER on HC3 and ChatGPT/Human Abstract datasets but low WER ($\sim$ 50\%) on the WikiText-2 dataset. Note that the WERs of \baserule{} are still 13\% lower than \sys{} even on the best-performing Human Abstract dataset.
}

%(1) Compared with \baserule{}, 64-bit \sys{} achieves 25\% better BERT-S and  similar average BLEU-4. Because \baserule{} changes synonyms without considering the overall semantics, which distorts the contents.  (2) The p-value \baserule{} provides is $\sim10^{-7}$ and equals an average z score of 5.19, whereas the average z-score \sys{} provides at 64-bit signature is 7.12. Therefore, \sys{} demonstrates better semantic preservation and watermarking strength compared with \baserule{}.

 %The table indicates \baserule{} maintains good semantic coherence by only changing the nouns and syntactic structure of the paragraph. However, the maximum watermarking strength (measured by p-value) \baserule{} can provide is $\sim10^{-7}$. Therefore, the maximum number of signature lengths that can be inserted by \baserule{} into paragraphs are smaller compared with \sys.
%\textbf{Comparison with \sys-segment: } 

\subsubsection{Watermarking Strength}\label{subsec:strength}
The watermarking strength, measured by z-score, quantitatively evaluates the statistical significance and robustness of the watermarks embedded within the content generated by the LLM. 
\revision{Based on Null Hypothesis~\cite{anderson2000null}, a z-score threshold of 1.64 corresponds to a 
p-value of less than 0.05, implying a significant presence of the watermarks in the content.
Following KGW~\cite{kirchenbauer2023watermark}, a higher z-score of 4 ($p = 3\times 10^{-5}$) indicates a greater alignment between the inserted and extracted signature bits. } This alignment serves as stronger evidence that the watermarks belong to the owner of the LLM.
% means more inserted and extracted signatures bits are aligned, thus providing stronger proof that the watermarks belong to the LLM proprietor.
We use the z-score from Section~\ref{subsec:wm_extract} to compute the watermarking strength for neural-based approach \sys{} and \baseend{}. The results are reported on the ChatGPT dataset with watermarking frameworks trained on HC3. The z-score of \baseinfer{}~\footnote{https://github.com/jwkirchenbauer/lm-watermarking} and \baseinferrb{}~\footnote{https://github.com/jthickstun/watermark} is calculated using their original implementation.  

%The adversary claims the watermark insertion is a coincidence by showing the z-score of a paragraph is below a certain threshold. We adopted the threshold from KGW~\cite{kirchenbauer2023watermark} and set the z-score below 4 as weak watermark insertions. 

%Following \baseinfer{}~\cite{kirchenbauer2023watermark},  we set a threshold of z-score=4 as a successful watermark insertion and a corresponding p-value of 3.16$\times10^{-5}$ can help prove the signatures are the exact ones inserted by LLM owner. This is denoted by the black dotted line in Figure~\ref{fig:strength}.

\begin{figure}[H]
  \vspace{-10pt}
    \begin{minipage}[c]{0.55\columnwidth}
\includegraphics[width=\textwidth]{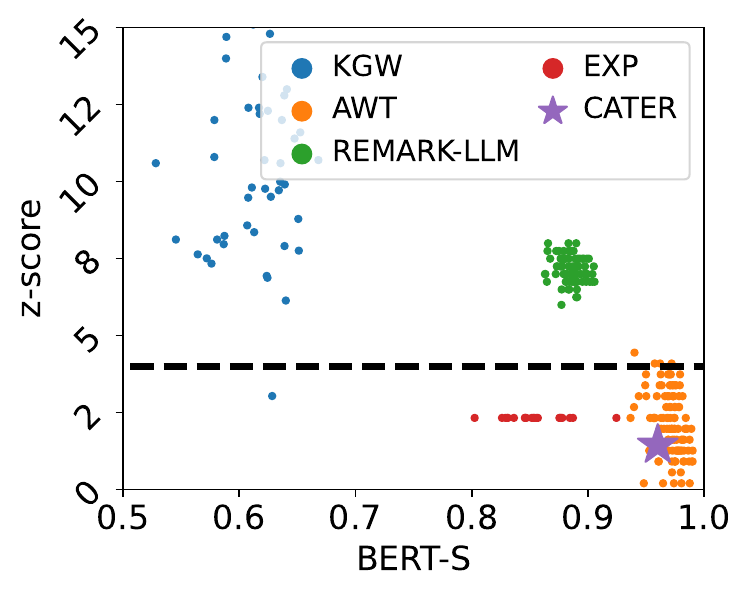}
    \end{minipage}
    \hfill
    \begin{minipage}[c]{0.44\columnwidth}
        \caption{Watermarking strength and semantic preservation comparison of different watermarking frameworks. The threshold for a strong watermark insertion is a z-score of 4, represented as the black dotted line. }\label{fig:strength}   
    \end{minipage}
    \vspace{-10pt}
\end{figure}

From Figure~\ref{fig:strength}, we find that (1) \baseend{}, \baseinferrb{}, and \baserule{} preserve the semantic integrity but fail to provide sufficient watermark strength for long text sequences. This makes these two approaches less suitable for real-world text ownership proof, where the adversary may argue the watermark insertion is coincidental. (2) While  \baseinfer{} effectively embeds watermarks to LLM-generated content, there is a noticeable semantic distortion and greatly degrading LLM quality. (3) \sys{} demonstrates the merits of the aforementioned approaches and ensures both semantic fidelity and watermarking strength. The average z-score of \sys{} is 
7.12, corresponding to an average one-side p-value of 5.4$\times 10^{-13}$. This provides adequate watermarking proof for a text sequence with 640 tokens. 
When operating LLMs like GPT-4  at their zenith, the generated contents will be a maximum of 8.2k tokens. \sys{} can provide a z-score of 25.20, corresponding to a one-tail p-value of smaller than 1.59$\times 10^{-130}$, all while maintaining semantic integrity.  As a result, \sys{} gives ample watermark strength to LLM proprietors. 

%3) \sys{} demonstrates the advantages of both methods and exhibits good semantic preservation and strong watermarking strength. The average z-score of the data samples is 7.12, corresponding to an average one-side p-value of 5.4$\times 10^{-13}$, which provides adequate and strong watermarking proof for a paragraph with 640 tokens. The protection is even stronger if LLM like GPT-4 generates contents at its full capacity and produces over 8.2k tokens. The z-score \sys{} can provide is 25.20, which equals a one-tail p-value of smaller than 1.59$\times 10^{-130}$. As a result, \sys{} gives ample watermark protection to LLM owners. 

%As a watermarking framework agnostic to the LLM architectures and data sources, \sys{} also provides real LLM systems with strong watermark signature verification. 

%\begin{figure}[!h]
%    \centering
%    \includegraphics[width= 0.8\columnwidth]{figs/bert_score_vs_p_value.pdf}
%    \vspace{-10pt}
%    \caption{Watermarking strength and semantic preservation comparison of different watermarking frameworks. The threshold for a strong watermark insertion is a z-score of 4, represented as the black dotted line. }
%\vspace{-10pt}
%    \label{fig:strength}
%\end{figure}

\subsubsection{Watermarking Efficiency} 
We measure the efficiency of inserting 8-bit watermarks into 100 samples of 80-token LLM-generated texts by reporting the average wall-clock time required in seconds and peak GPU memory utilization in GB. 
These results are summarized in Table~\ref{tab:efficient}.
For \sys{} and \baseend{}, we report the inference time for the watermarking framework to generate the watermarked texts. For \baseinfer{} and \baseinferrb{}, we report the time difference between the inference time with and without watermark insertion. 

The results indicate that inserting the same number of bit signatures into the LLM-generated texts, \baseinfer{} requires nearly 2 $\times$ more time compared with neural-based \sys{} and \baseend. The additional overhead stems from the green/red list splitting over the vocabulary for every next token prediction. For  \baseinferrb{}, which aims to improve the inference-time watermarking's semantic preservation, the time overhead becomes even larger, which takes 14$\times$ longer than \sys{} to watermark the same texts. This greatly constraints such approaches from being widely applied in real-world LLM APIs.  

%For \baserule, the overhead includes tagging Part-of-Speech (POS) for each word and finding noun synonyms with Word2Vec model~\cite{mikolov2013efficient}. The optimization for watermark insertion is unavailable upon our submission, and we did not include the time in Table~\ref{tab:efficient}. The time taken for \baserule{} insertion will be substantially longer if considering the optimizations.

For the computation resources, \baseinfer{} and \baseinferrb{} require 2.1$\times$ more GPU memory compared with \sys{} and 4.5$\times$ compared with \baseend. The required computation resources for neural-based \sys{} and \baseend{} are agnostic to the LLM type, whereas \baseinfer{} and \baseinferrb{} require more GPU memory as LLM size grows. Therefore, the neural-based approaches are more computationally efficient compared with inference-time frameworks. For the rule-based \baserule, the textual transformations are performed on the CPU, and the GPU memory consumption is 0.

\begin{table}[!ht]
  \centering
 \small
 \vspace{-10pt}
  \begin{tabular}{ccc}
    \toprule
    Methods &  Time (s) & Memory(GB) \\
    \midrule
   \sys & 1.21 & 5.83\\
   \baseend~\cite{abdelnabi21oakland} & 1.14 & 2.72 \\
   \baseinfer~\cite{kirchenbauer2023watermark}  &2.36 & 12.21 \\
   \baseinferrb~\cite{kuditipudi2023robust} & 17.50 & 12.21\\
   \baserule~\cite{he2022cater} & 1.02 & 0 \\
    \bottomrule  
  \end{tabular}
  \vspace{-5pt}
  \caption{Efficiency (time and memory) comparison among watermarking frameworks. \label{tab:efficient}}
 \vspace{-10pt}
\end{table}

\subsubsection{Watermarking Integrity}

\revision{We evaluate \sys's watermarking integrity by running its message decoding module on non-watermarked texts. Then, we calculate the WERs by comparing the decoded signature with the randomly generated encoding signature. 
We benchmark the WERs on the four datasets' test set and report the results in Table~\ref{tab:integrity}.  The close to 50\% WER on non-watermarked texts demonstrates \sys's integrity.}

% & \begin{tabular}[c]{@{}c@{}}ChatGPT\\Abstract\end{tabular} & \begin{tabular}[c]{@{}c@{}}Human\\Abstract\end{tabular}Passing non-watermarked text through the message decoding module results in random bit-strings
%Based on probability theory, the watermark extraction rate of such an output against a given randomly generated secret is 50\%. As a result, the 50\% WER on non-watermarked texts demonstrates \sys's integrity.
\begin{table}[!ht]
  \centering

 \vspace{-5pt}
  \resizebox{\columnwidth}{!}{
  \begin{tabular}{ccccc}
    \toprule
    Datasets &  HC3 & WikiText-2 & \begin{tabular}[c]{@{}c@{}}ChatGPT Abstract\end{tabular} & \begin{tabular}[c]{@{}c@{}}Human Abstract\end{tabular} \\
    \midrule
   WER & 49.9\% & 49.9\% & 50.0\% & 50.1\%\\
    \bottomrule  
  \end{tabular}}
  \vspace{-10pt}
  \caption{Watermarking integrity on different datasets. 
  \vspace{-20pt}
  \label{tab:integrity}}
\end{table}

\begin{table*}[!ht]
\small
\centering
\resizebox{\textwidth}{!}{%
\begin{tabular}{p{9.5cm}p{9.5cm}}
\toprule
Original Text &  Watermarked Text  \\ \midrule
It can be hard \hl{to} explain \hl{it} in simple terms. But I'll \hl{do} my best! During \hl{inflation}, the universe expands at an incredibly fast rate. But it's important to \hl{note} that this expansion is not \hl{like} the movement of objects through space. & 
It can be hard \hl{directly} explain \hl{exactly} in simple terms. But I'll \hl{try} my best! During \hl{time}, the universe expands at an \hl{infinite} incredibly fast rate. But it's important to \hl{understand} that this expansion is not \hl{about} the movement of objects through space \hl{itself}. \\\hline
In the context of \hl{financial investments}, "headwinds" refer \hl{to} negative factors that can potentially \hl{hinder} the \hl{performance} of an investment. These \hl{may} include economic \hl{conditions}, regulatory \hl{changes}, market \hl{trends}, or other external factors that \hl{can} work against the investment.  & 
In the context of \hl{stocks investing}, "headwinds" refer \hl{for} negative factors that can potentially \hl{impact} the \hl{value} of an investment. These \hl{factors} include economic \hl{impacts}, regulatory \hl{issues}, market \hl{conditions}, and other external factors that work against the investment.\\\toprule
The paper \hl{discusses} Colombeau’s \hl{generalized} function on arbitrary manifolds. We first \hl{define} the space of Colombeau’s generalized functions by \hl{quotienting} out \hl{by a suitable} ideal endowed with a ring structure. & The paper \hl{introduced} Colombeau’s \hl{generalization} function on arbitrary manifolds. We first \hl{study} the space of Colombeau’s generalized functions by \hl{particle} out ideal endowed with a ring structure.\\\hline
This \hl{paper} presents a novel \hl{methodology} for constructing super throats using non trivial scalar fields.
By \hl{introducing} these fields, we are able to \hl{achieve unprecedented} control over the dynamics of the throats.&  This \hl{research} presents a novel \hl{approach} for constructing super throats through non trivial scalar fields. By \hl{utilizing} these fields, we are able to \hl{obtain precise} control over the dynamics of the throats.\\  
\bottomrule
\end{tabular}}
\vspace{-10pt}
\caption{Watermarked Text Examples. All of the watermarked texts achieve 100\% WER. The first two examples are randomly taken from the HC3 test set, and the last two are randomly taken from the ChatGPT Abstract test set. The edited words are highlighted in \hl{yellow}. \label{tab:visual}}
\vspace{-10pt}
\end{table*}

\subsubsection{Watermarking Different LLM Architectures}
We have shown \sys{} trained on the ChatGPT-generated HC3 dataset can successfully watermark human written texts and GPT-3.5 Turbo-generated ChatGPT Abstract in Section~\ref{tab:effective}. We also investigate whether trained \sys{} can watermark unseen texts generated by other LLM families in Table~\ref{tab:llm-family}. We use 2k instruction prompts from Alpaca Dataset~\cite{alpaca} and watermark responses from three state-of-the-art open-source LLMs: OPT~\cite{zhang2022opt}, OpenOrca~\cite{mukherjee2023orca}, and LLaMA-2~\cite{touvron2023llama}. Those models are trained with different architectures and different training datasets. The prompts for generating those responses do not overlap with \sys's training dataset HC3. 

The watermarking is performed for long text sequences by using the instruction prompts as input and limiting the LLMs to predict the next 640 tokens. \sys{} insert 64 bits into their generated texts and report the watermarking performance in Table~\ref{tab:llm-family}. We find \sys{} can successfully insert and extract watermarks from texts generated by different LLM architectures. \sys{} also keeps high-quality semantic preservation and coherence. Our results indicate that once \sys{} is trained on the large LLM-generated corpus like HC3, the trained \sys{} is agnostic to the  LLM architectures and data sources. \sys{} can thereby be used in real-world LLM watermarking scenarios. 

\begin{table}[!ht]
\centering
\small
\vspace{-10pt}
\begin{tabular}{cccc}
\toprule
LLMs & WER(\%) & BERT-S &BLEU-4 \\ \midrule
OPT-2.7B~\cite{zhang2022opt}& 93.42 & 0.91 &0.34\\  
OpenOrca-7B~\cite{mukherjee2023orca} & 93.70 & 0.92 &0.35  \\  
LLaMA-2-7B~\cite{touvron2023llama} & 91.18 & 0.91 &0.39 \\  
\bottomrule
\end{tabular}
\caption{\sys{} performance in watermarking texts generated by different LLM architectures. \label{tab:llm-family}}
 \vspace{-20pt}
\end{table}

\subsubsection{Watermarking Examples} 

We present the original LLM-generated and watermarked texts in Table~\ref{tab:visual} and highlight their differences. The first two texts are taken from the HC3 test set, and the last two are taken from the ChatGPT Abstract test set.  All texts are embedded with 8-bit message signatures using \sys{} trained on HC3's training set at the segment level. For all the watermarked messages presented in Table~\ref{tab:visual}, the watermarks are successfully extracted with 100\% WER. We include more watermarking examples in the Appendix~\ref{sec:append_wm_example}.

From Table~\ref{tab:visual}, we find that the watermarked texts are readable and possess the same semantics as the original texts. \sys{} not only learns to replace the word with their synonyms but also adds/deletes words or replaces other non-synonyms to ensure coherence and preserve the semantic structure. For example, in the first example, ``itself'' is added after ``space''; in the second example, ``may'' is replaced by ``factors''; and in the third example, ``by a suitable'' is deleted. Those changes do not affect the overall quality of the watermarked texts but help to accommodate longer watermark signature sequences.

%\begin{table*}[!h]
%  \centering
%  \small
%  \begin{tabular}{ccccc}
%    \toprule
%    Dataset & {\begin{tabular}[c]{@{}c@{}}Dist.\\ Gen. (ms)\end{tabular}}   & {\begin{tabular}[c]{@{}c@{}}Beam\\ Search (s)\end{tabular}} &  {\begin{tabular}[c]{@{}c@{}}Total\\ (s)\end{tabular}}  & {\begin{tabular}[c]{@{}c@{}}Memory\\  (GB)\end{tabular}}  \\
%    \midrule
%    HC3 &0.03 &1.98 &2.01 & 5.92 \\
%    ChatGPT Abstract & 0.03 & 1.53& 1.56 & 5.84 \\
%    Human Abstract & 0.03 &1.93 & 1.96 & 5.89\\
%    WikiText-2 & 0.03 & 1.18& 1.21 & 5.83\\
%    \bottomrule  
%  \end{tabular}
%  \caption{\sys{} watermark insertion overhead, regarding the time and computation resource. \label{tab:overhead}}
%\end{table*}

%In this subsection, we measure \sys's transferability by training it on the ChatGPT dataset and watermarking the validation texts of the datasets in Table~\ref{tab:cross}. The inserted message length is 16-bit for all data, where samples with token sizes smaller than 496 are padded to the maximum length.

%From here, we can find (1) \sys{} exhibits good transferability among datasets. 

\subsection{Ablation Studies}
In this section, we study how different hyperparameter choices affect the performance of \sys{} during both training and inference time. 

\subsubsection{Training-time Ablation Study}
In this section, we study how different training coefficients affect the watermarking performance. The results in this section are all trained on the HC3 dataset with 8 bits inserted in an 80-token segment. The inference hyperparameters follow the default settings from Appendix~\ref{sec:parameter_define} and are reported on the ChatGPT Abstract dataset to avoid overfitting.

\noindent\textbf{Effectiveness of Watermark Backbones} We use the default training settings but change the Seq2Seq Model's backbone from T5-base to T5-small and T5-large. For the two replaced models, T5-small has 8 attention heads, 6 layers, and 512 feedforward dimensions, whereas T5-large has 16 attention heads, 24 layers, and 1024 feedforward dimensions.  

From Table~\ref{tab:backbone}, as the backbone Seq2Seq model becomes larger, \sys{} gets better text coherence and better watermark extraction rates. This is because larger backbone models have more parameters to learn how to add watermarks in the feature space and achieve better performance.  This improvement is significant when the model is switched from T5-small to T5-base, and is marginal when the model is changed from T5-base to T5-large. 

\begin{table}[!ht]
 \small
\centering
\begin{tabular}{cccc}
\toprule
Backbone & WER(\%) & BERT-S &BLEU-4 \\ \midrule
T5-small & 91.60& 0.89 & 0.25  \\  
T5-base& 93.53 & 0.91 &0.29  \\  
T5-large & 93.77 & 0.91 &0.32  \\  
\bottomrule
\end{tabular}
\vspace{-5pt}
\caption{The effectiveness of different model backbones in \sys{} performance. \label{tab:backbone}}
 \vspace{-10pt}
\end{table}

\noindent\textbf{Effectiveness of Masking Percentages and Gumbel Temperatures}
We compare how different masking percentages and the Gumbel-Softmax temperatures affect the overall text coherence and watermark extraction accuracy. The results are summarized in Table~\ref{tab:hyper1}, where we change the input masking percentage from 30\% to 70\% and the Gumbel-Softmax temperatures from 0.1 to 0.5.

We find that: (1) As the masking percentage becomes larger, at the same Gumbel-Softmax temperature, the coherence between LLM-generated texts and the original texts worsens, whereas the watermarking extraction rates become higher. Given that \sys{} has more masked space, it introduces more alterations to the original texts, thereby can accommodate more signatures.
(2) As the Gumbel-Softmax temperature becomes larger, at the same masking percentage, the coherence of the texts becomes worse, but more watermark signature bits are extracted from the watermarked texts. However, if the temperature rises to 0.5, \sys{} fails to learn the one-hot encoding from Seq2Seq output and achieves worse watermark extractions. 
Therefore, we trade off the semantic preservation and the watermarking extraction rates during  \sys{} training. By default, we choose a masking percentage of 50\% and a Gumbel-Softmax temperature of 0.3.

\begin{table}[!ht]
 \small
\centering
\begin{tabular}{ccccc}
\toprule
Masking & Temp & WER(\%) & BERT-S &  BLEU-4 \\ \midrule
\multirow{3}{*}{30\%} & 0.1 &  89.78 & 0.91 & 0.34\\  
& 0.3 &  89.97  & 0.91 & 0.31\\  
& 0.5 & 83.09 & 0.91 & 0.31\\  
 \midrule
\multirow{3}{*}{50\%} & 0.1 & 91.18 & 0.91 & 0.31  \\  
& 0.3 & 93.53 & 0.91 & 0.29 \\  
& 0.5 & 85.43 & 0.90 & 0.29\\ 
 \midrule
\multirow{3}{*}{70\%} & 0.1 & 94.74 & 0.89 & 0.26 \\  
& 0.3 & 94.98 & 0.87 & 0.17 \\  
& 0.5 & 87.30 & 0.86 & 0.17 \\ 
\bottomrule
\end{tabular}
\vspace{-5pt}
\caption{The effectiveness of different masking percentages and gumbel noises in \sys{} performance.\label{tab:hyper1}}
 \vspace{-10pt}
\end{table}
%The results in this subsection are all trained on HC3 dataset with 8 bits inserted per 80 token segment. 

\noindent\textbf{Effectiveness of Loss Coefficients} We include the watermarking performance under different loss coefficients in Table~\ref{tab:train_coeff}. The $w_1$ semantic loss coefficient increases from 0.3 to 0.7, whereas the $w_2$ message recovery loss coefficient decreases from 0.7 to 0.3. From here, we find that larger $w_1$ indicates better semantic coherence between the original and watermarked texts and a larger $w_2$ results in higher watermark extraction rates. Therefore, we choose $w_1=0.5$ and $w_2=0.5$ to obtain a balance between the text semantic coherence and watermark extractions. 

\begin{table}[t]
 \small
\centering
\begin{tabular}{ccccc}
\toprule
$w_1$ & $w_2$ &WER(\%) & BERT-S &  BLEU-4  \\ \midrule
0.3 & 0.7 &  94.78 &  0.89 & 0.25  \\  
0.5 & 0.5 & 93.53 & 0.91 & 0.29 \\  
0.7 & 0.3 & 91.76 & 0.92 & 0.33\\  
\bottomrule
\end{tabular}
 \vspace{-5pt}
\caption{The effectiveness of different loss coefficients in \sys{} performance.\label{tab:train_coeff}}
 \vspace{-10pt}
\end{table}

\subsubsection{Inference-time Ablation Study}
In this section, we study how different coefficients in the inference time affect the \sys{} performance. The training hyperparameters follow the default settings from Appendix~\ref{sec:parameter_define}. We change the token masking percentage in Algorithm~\ref{alg:insertion}. The results are summarized in Table~\ref{tab:infer_hyper}, where we change the masking input percentage from 0.3 to 0.7.

From here, we find that as the masking percentage grows, \sys{} tends to get better WER. However, the coherence and semantic preservation of the texts are compromised, as reflected by the lower BLEU-4 and BERT-S. Therefore, we choose the masking percentage during inference to be 50\% to ensure semantic preservation as well as high watermark extraction rates. 

\begin{table}[!ht]
 \small
\centering
\begin{tabular}{cccc}
\toprule
Masking &WER(\%) & BERT-S &  BLEU-4  \\ \midrule
30\%& 91.73 & 0.93 & 0.45 \\  
50\%& 93.53 & 0.91 & 0.29 \\  
70\%& 96.22 & 0.86 & 0.12 \\  
\bottomrule
\end{tabular}
\caption{The effectiveness of different inference hyperparameters in \sys{} 
performance.\label{tab:infer_hyper}}
 \vspace{-10pt}
\end{table}

%\subsubsection{Watermark Capacity} 

%\sys's watermarking capacity is displayed in Figure~\ref{fig:capacity}, where with the given ChatGPT dataset, we increase the inserted watermarking bits from 4 bits to 32 bits. From here, we can find a trade-off between the maximum bits that can be accommodated and the watermark extraction rates. The cut-off threshold is 16 bits per 496 tokens. 

%\begin{figure}[!h]
%    \centering
%    \includegraphics[width= 0.6\columnwidth]{figs/wer.pdf}
%    \caption{Watermark performance by accommodating different message bit lengths. }
%    \label{fig:capacity}
%\end{figure}

\subsection{Attack Evaluations}\label{subsec:attack}

\revision{In this section, we evaluate the watermarking frameworks' \textbf{Robustness} under watermark removal attacks and \textbf{Undetectability} under watermark detection attacks.}

\begin{figure*}[!ht]
    \centering
    \subfloat[Text Deletion Attack]{\label{f:delete} \includegraphics[width=0.2\linewidth]{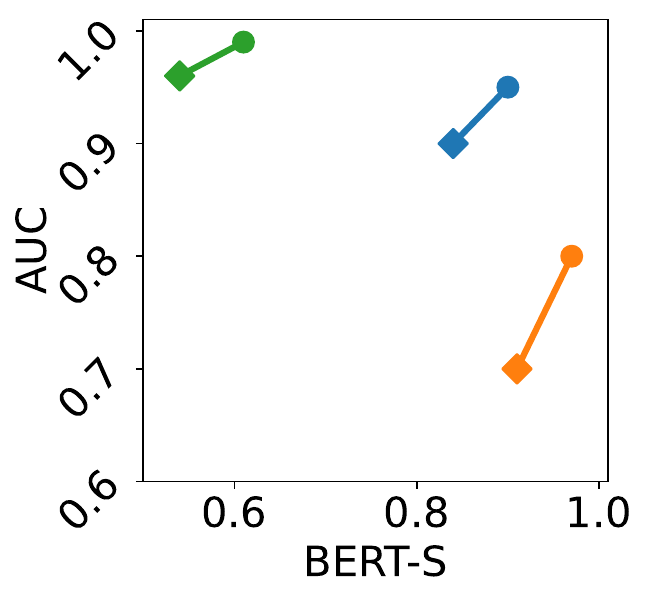}}
    \subfloat[Text Addition Attack]{\label{f:addition} \includegraphics[width=0.2\linewidth]{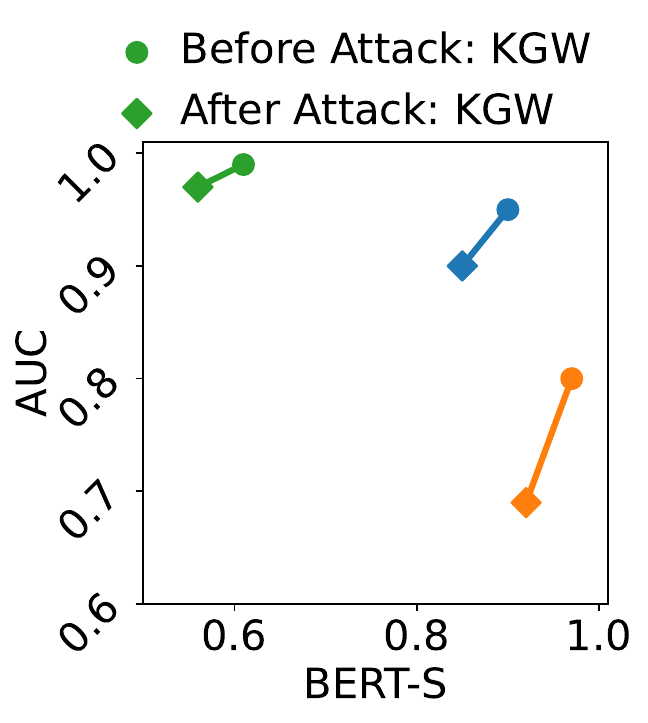}}  
    \subfloat[Text Replacement Attack]{\label{f:replacement} \includegraphics[width=0.2\linewidth]{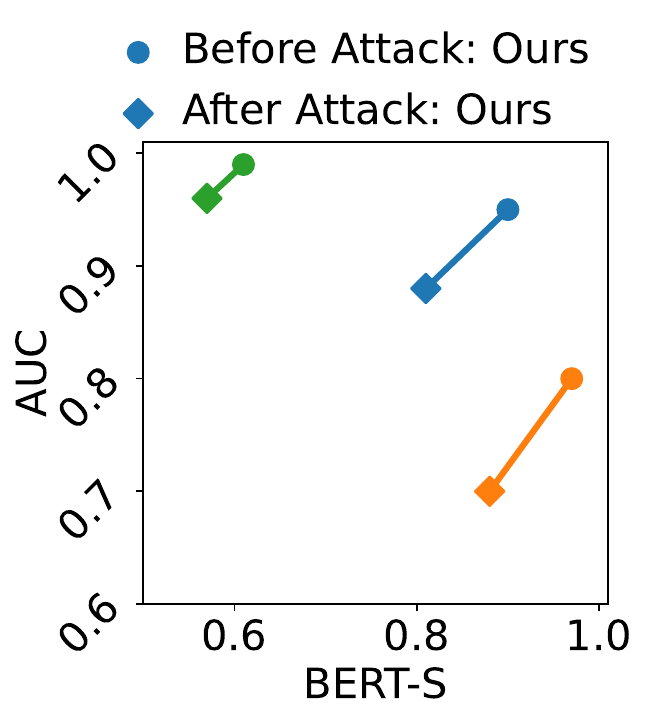}}
    \subfloat[Text Rephrase Attack]{\label{f:rephrase} 
    \includegraphics[width=0.2\linewidth]{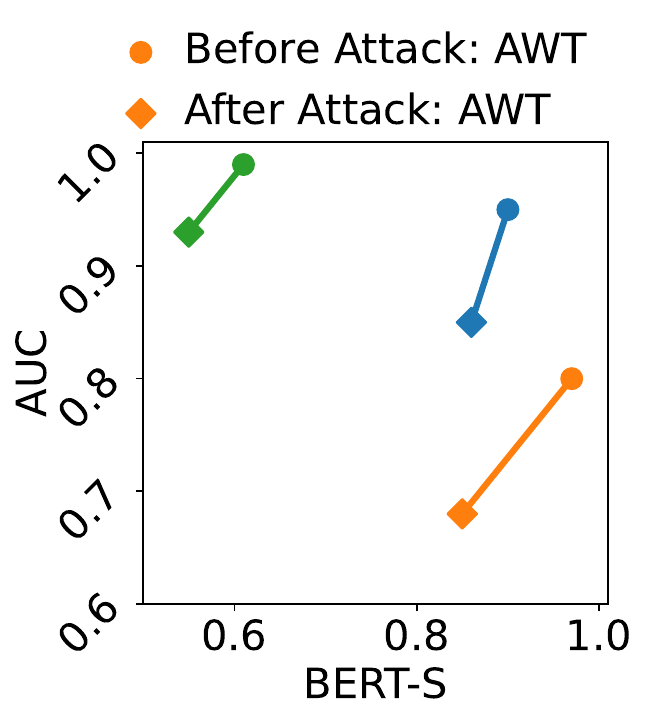}}  
    \subfloat[Re-watermark Attack]{\label{f:rewateramrk} \includegraphics[width=0.2\linewidth]{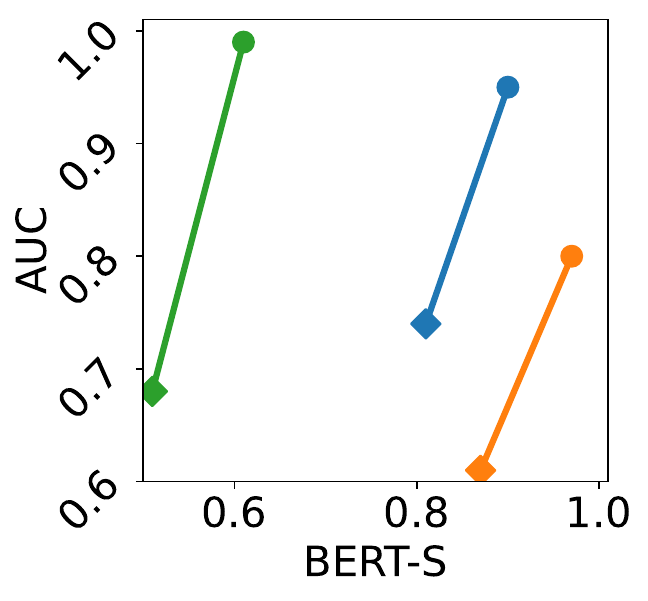}} 
    \caption{Watermarking performance under different attacks, including watermark extraction measured by AUC and semantic coherence measured by BERT-S. The attacks are performed on the ChatGPT Abstract dataset with frameworks trained on the HC3 dataset. \baseinfer~\cite{kirchenbauer2023watermark} is the inference-time watermarking framework. \baseend~\cite{abdelnabi21oakland} is the neural-based watermark framework. % The bullet marker $\bullet$ denotes watermarking performance without attack, whereas the square marker $\blacksquare$ denotes watermarking performance after attack. 
    From left to right, we study text edit attacks (deletion, addition, and replacement), text rephrase attacks and re-watermark attacks.}
 \vspace{-5pt}
    \label{fig:ROC}
\end{figure*}

\subsubsection{Watermark Removal Attacks}
In this section, we demonstrate the resiliency of  \sys{} under watermark removal attacks. %We include more analysis on whether the watermarked texts can be detected using statistical or machine learning-based approaches in  Appendix~\ref{sec:append_wm_attack}.
The adversary attempts to remove the signatures in the watermarked texts by performing the following attack: (1) \textbf{Text Deletion Attack}: randomly delete words in the watermarked texts with a probability of 6\%; (2) \textbf{Text Addition Attack}: randomly add words in the watermarked texts with a probability of 6\%; (3) \textbf{Text Replacement Attack}: randomly replace words with their synonyms by leveraging Word2Vec~\cite{mikolov2013efficient} pre-trained on Google News corpus~\cite{mikolov2013distributed} (3 billion running words); (4) \textbf{Text Rephrase Attack}: using open-source NLP tool T5-large~\cite{raffel2020exploring} to rephrase the watermarked texts with prompt ``paraphrase the texts:" and accept the rephrase if the BERT-S between the watermarked and rephrased sentence is higher than 0.85; and (5) \textbf{Re-watermark Attack}: training a local version of \sys{} with WikiText-2 dataset, and re-watermark the texts with new signatures. We report \sys's performance under the aforementioned attacks using \sys{} trained on the HC3 dataset and report the attack performance on the ChatGPT dataset. % The ROC curves for different watermarking frameworks are shown in Figure~\ref{fig:ROC}.

The adversary's goal is to remove the signature and 
maintain the semantics and readability of the resultant texts. Therefore, we do not consider the emojis or special characters attack proposed in ~\baseinfer~\cite{kirchenbauer2023watermark}, which adds additional symbols into the watermarked contents to compromise the overall text readability. Furthermore, the adversary uses LLM-generated texts to reduce the time for batch-generating spam content. Therefore, we do not consider the scenario where extra human labor is involved in the attack to rewrite the texts for watermark removal, which deviates from the adversary's initial objectives.

\noindent\textbf{Text Deletion Attack} From Figure~\ref{f:delete}, we find that (1) \sys{} achieves an AUC of 0.90 under Text Deletion Attack. This demonstrates that 90\% of watermarking and non-watermarking texts can be successfully classified, providing sufficient ownership proof for LLM proprietors. (2) While \baseinfer{} achieves 0.96 AUC, the semantic preservation evaluated by BERT-S even before the attack is 0.61. Therefore, \baseinfer{} failed to provide effective watermarking to LLM-generated content. (3) \baseend{} successfully maintains the semantics with BERT-S larger than 0.90, but the AUC drops to 0.70, demonstrating worse resilience towards text deletion attacks.

\noindent\textbf{Text Addition Attack} From Figure~\ref{f:addition}, we find (1) \sys{} maintains high watermark extraction rates, yielding an average AUC of 0.90 and average BERT-S of 0.90 before attack.  (2) Similar to the text deletion attack, \baseend's average AUC degrades to 0.69, which is 23\% lower compared with \sys{} and exhibits worse resilience toward text addition attack. (3) While \baseinfer{} also demonstrates an average AUC of 0.97 after attacks, it fails to maintain semantic coherence in generated texts before and after attacks. This indicates that by including malicious transformations during training, \sys{} becomes robust to unseen threats and maintains semantic coherence during watermarking.
 
\noindent\textbf{Text Replacement Attack} The replacement attack poses a more significant threat to watermark extraction than the prior two approaches, as depicted in Figure~\ref{f:replacement}. (1) Firstly, \baseend{} and \sys{} watermarking extraction relies on combinations of textual tokens to decode the inserted messages from their respective feature space. By replacing words with synonyms, some of the features used for watermark extraction are compromised, thus resulting in worse AUC. However, the replacement attack does come with a cost of 4-5\% BERT-S degradations. 
(2) \sys{} maintains an AUC of 0.88 with a 0.90 BERT-S score, demonstrating its resilience under such attacks. In contrast, \baseend's AUC degrades to 0.70, and \baseinfer's BERT-S degrades from 0.61 to 0.57. This demonstrates that \sys{} maintains a good trade-off between robustness and semantic preservation of watermarked content.

\noindent\textbf{Text Rephrase Attack} Figure~\ref{f:rephrase} demonstrates that text rephrase attack results in stronger WER degradation compared to text editing attacks. 
(1) Our findings show that \sys{} remains resistant to rephrase attacks and consistently maintains an AUC of 0.85. This translates to an average z-score of 5.6 for long text sequences, highlighting the robustness of \sys{} in establishing ownership proof.
% (1) Nevertheless, we find \sys{} is resilient to rephrase attacks and maintains an AUC of 0.85, which equals an average z-score of 5.6 for long text sequences, demonstrating \sys's resilience for ownership proof. 
(2) Compared to baselines, \baseinfer's maintains an AUC of 0.92, but its BERT-S drops to 0.61 even before the attack. (3) Conversely, while \baseend{} preserves semantic information before attacks, its AUC drops to 0.68 after the attack. 

\noindent\textbf{Re-watermark Attack} We highlight re-watermarking performance in Figure~\ref{f:rewateramrk}. Re-watermarking degrades the WER by rephrasing the text with \sys{} and tries to break the text features used for watermark verification.  Therefore, the WER for all three methods significantly degrade. 
We find that \sys{} maintains an AUC of 0.74, whereas \baseinfer{} degrades to 0.68 and \baseend{} degrades to 0.61, demonstrating \sys's robustness to re-watermarking attacks.

\subsubsection{\revision{Watermark Detection Attacks}}
\label{sec:append_wm_attack}

In this section, we include more security evaluations on \sys. Instead of removing the watermark, the adversary intends to know if watermarks are presented in specific texts.  One of the most straightforward ways is to use the message decoding module to detect if the watermarks can be successfully extracted. However, when the adversary is a malicious end-user, he/she will not have access to \sys's trained components. To detect the watermark presence, the adversary uses (1) statistical analysis of watermarked texts and (2) machine-learning models to classify watermarked and non-watermarked texts.

\begin{figure}[!ht]
\vspace{-10pt}
    \centering
    \includegraphics[width= \columnwidth]{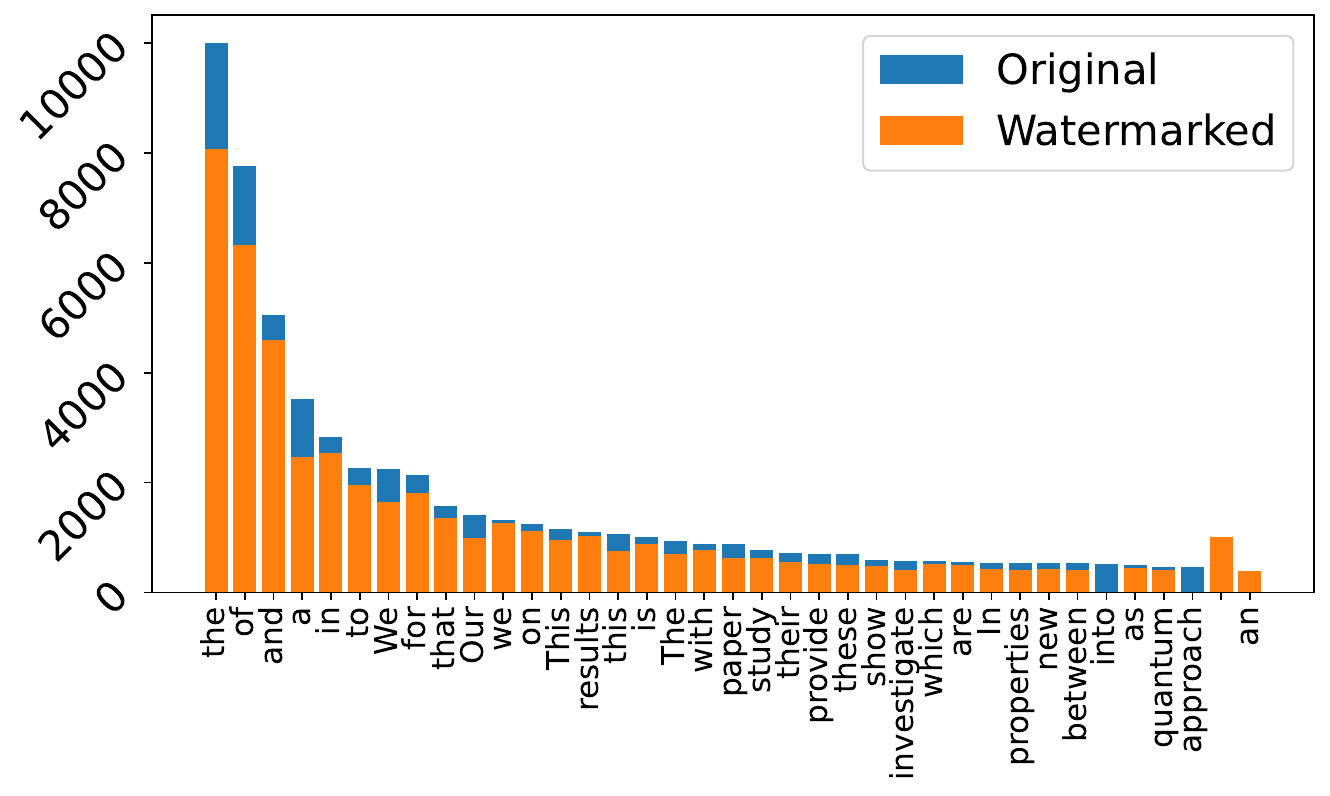}
    \caption{Word frequency distribution of original LLM-generated texts and corresponding watermarked texts. }
    \label{fig:freq_word}
    \vspace{-10pt}
\end{figure}

\noindent\textbf{Statistical analysis} We show top-word distribution from the original LLM-generated texts and the watermarked texts in Figure~\ref{fig:freq_word}. The texts are taken from the ChatGPT abstract dataset with \sys{} train on HC3. From here, we can find the distributions are close, meaning the adversaries cannot distinguish if texts are watermarked or not by simply analyzing the word frequency distributions. 

\noindent\textbf{Machine-learning classification} We use machine learning models to classify 2k samples with 1k watermarked and 1k LLM-generated texts. Those texts are non-paired, meaning the watermarked version of the LLM-generated texts is not shown in the training dataset. This is a common setting in which the adversary cannot access original LLM-generated texts. The adversary can only request the LLM API for watermarked texts and search unwatermarked texts online. 
We use another 500 LLM-generated texts and their corresponding watermarked texts to form the test set. The performance of different machine learning models is shown in Table~\ref{tab:ml_classification}. We train a 3-layer transformer~\cite{wolf-etal-2020-transformers} (8 heads with 512 dimensions), a BERT-base~\cite{devlin2018bert}, and a BERT-large~\cite{devlin2018bert} on the training dataset and report the detection performance on the test dataset. 

From Table~\ref{tab:ml_classification}, we can find that the accuracy is 50\%, and the F1-Score is 0 for all three models. This suggests that machine-learning models cannot confidently classify whether the texts are watermarked or not. Therefore, the watermark detection attacks fail to detect whether the LLM-generated contents are watermarked or not. 

\begin{table}[t]
 \small     
    \centering
    \begin{tabular}{c|c|c}
    \toprule
       Model  &  Acc. (\%) & F1-Score\\
     \midrule
       Transformer~\cite{wolf-etal-2020-transformers}  & 50.00  & 0\\\
       BERT-base~\cite{devlin2018bert} & 50.00  & 0\\
       BERT-large~\cite{devlin2018bert} &  50.00  & 0\\
    \bottomrule  
    \end{tabular}
    \vspace{-5pt}
    \caption{Classification performance on watermarked and non-watermarked texts. This shows the watermark detection attacks failed to detect whether the LLM-generated contents are watermarked. }
     \vspace{-15pt}
    \label{tab:ml_classification}
\end{table}

\subsection{Overall Evaluations}

\begin{figure}[H]
  \vspace{-10pt}
    \begin{minipage}[c]{0.55\columnwidth}
\includegraphics[width=\textwidth]{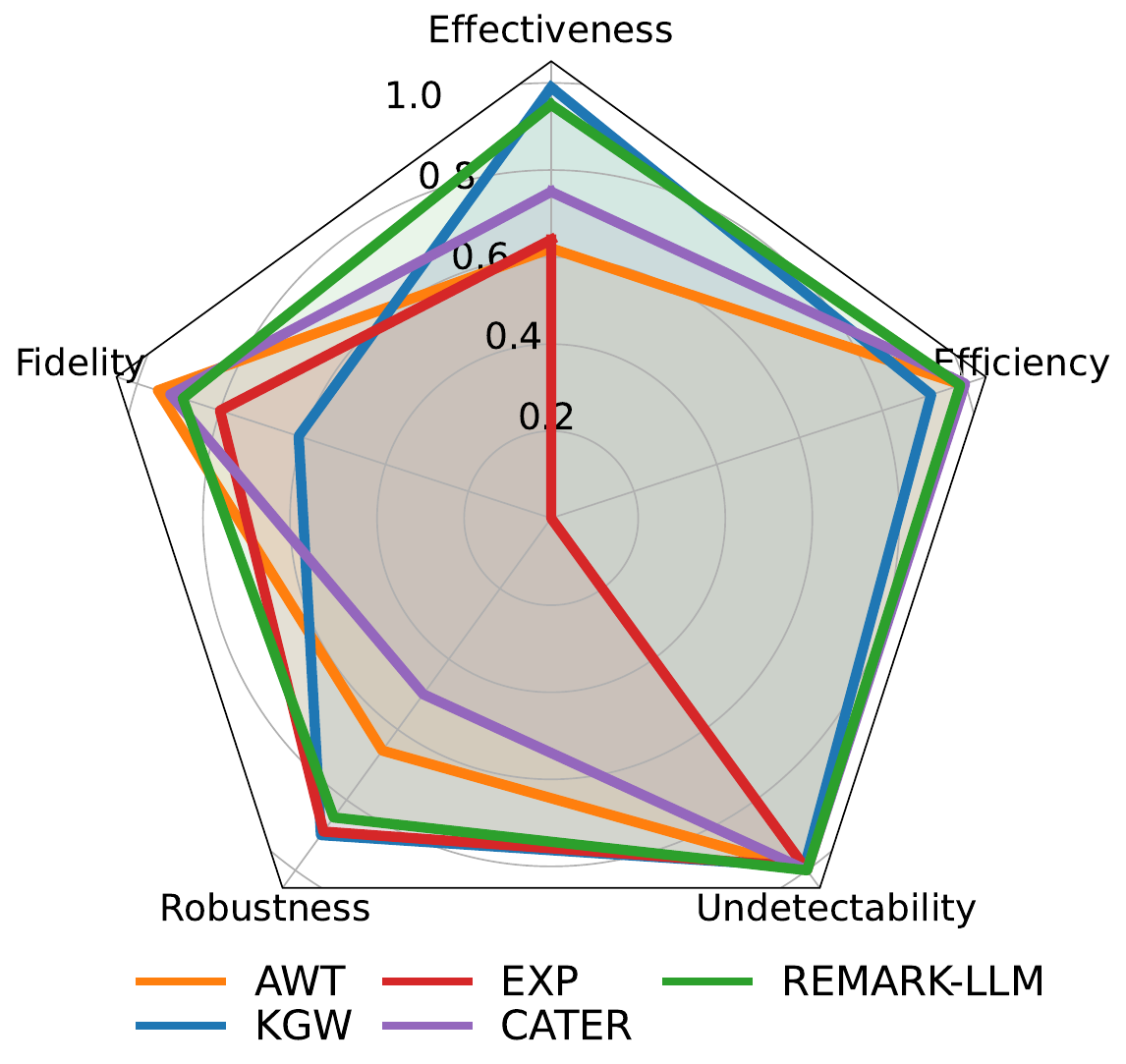}
    \end{minipage}
    \hfill
    \begin{minipage}[c]{0.44\columnwidth}
        \caption{Overall evaluations of different frameworks. \sys{} maintains good performance over all the criteria for watermarking, whereas other baselines demonstrate sensitive trade-offs. }\label{fig:overall_eval}   
    \end{minipage}
    \vspace{-22pt}
\end{figure}

\revision{In the previous experiments, we evaluate different watermarking frameworks' performance in terms of (i) effectiveness, (ii) fidelity, (iii) efficiency, (iv) robustness, and (v) undetectability. Their corresponding results are summarized in Figure~\ref{fig:overall_eval}. 
The effectiveness is measured by WER, fidelity is measured by BERT-S, and the robustness is measured by the average AUC after removal attacks on the ChatGPT dataset as in Table~\ref{tab:effective-paragraph} and Figure~\ref{fig:ROC}.
Undetectability is measured by 1-F1-score with ML-based classification and efficiency is the normalization of average WM insertion time in Table~\ref{tab:efficient}.
Higher numbers are desirable for all metrics.
We can observe that prior works demonstrate very sensitive trade-offs among all the criteria. However, \sys{} maintains high performance over all the requirements, making it an ideal toolkit for real-world watermarking applications. }

% The performance is benchmarked on the WikiText-2 dataset, where effectiveness is measured by WER, fidelity is measured by BERT-S, and robustness is measured by the average AUC after removal attacks.

%\begin{table}[h]
%\centering
% \resizebox{\columnwidth}{!}{
%\begin{tabular}{cccccc}
%\toprule
%Criteria & \sys & \baseend~\cite{abdelnabi21oakland} & \baseinfer~\cite{kirchenbauer2023watermark} & \baseinferrb~\cite{kuditipudi2023robust} & \baserule~\cite{he2022cater} \\ \midrule
%WM Strength & \cmark & \xmark  & \cmark &  \xmark & \xmark\\  
%Efficiency&  \cmark  &  \cmark  &   \cmark & \xmark & \cmark \\  
%Robustness & \cmark  &  \xmark  &  \cmark & \cmark & \xmark\\  
%Fidelity & \cmark & \cmark &  \xmark & \cmark & \cmark\\  
%\bottomrule
%\end{tabular}
%}
%\vspace{-10pt}
%\caption{Overall evaluation of the watermarking frameworks.\label{tab:compare}}
%\vspace{-10pt}
%\end{table}

%\section{Discussion}
%\input{files/6_discussion}

\vspace{-0.1cm}
\section{Conclusion and Future Work}
\label{sec:conclusion}
We present \sys{}, a robust and efficient framework for watermarking text contents generated by LLMs. \sys{} trains message encoding, reparameterization, and message decoding modules jointly, with the aim to accommodate the watermark signatures into the LLM-generated content while ensuring semantic coherence. 
The message encoding module facilitates the embedding of watermarks into LLM-generated content, while the decoding module extracts messages from these watermarked texts.
Comprehensive experiments have demonstrated that \sys{} can embed up to 64 binary bits within 640-token texts without compromising semantics and coherence. Furthermore, \sys{} exhibits transferability in watermarking unseen data sources and LLM architectures without additional fine-tuning and showcases resilience against various watermark detection and removal attacks. 

%\revision{For future work, we will explore the transferability of \sys{} toward watermarking more domain-specific data, such as code snippets and medical data, by incorporating additional training objectives. Besides, we are also investigating adaptive approaches for verifying the watermarked contents within the documents. }

\revision{We recommend future work in exploring the transferability of \sys{} towards watermarking more domain-specific data, such as code snippets and medical data, by incorporating additional training objectives. We also recommend future work to investigate adaptive approaches for verifying watermarked contents within the documents.}

\bibliographystyle{plain}
\bibliography{bib}

\begin{thebibliography}{10}

\bibitem{abdelnabi21oakland}
Sahar Abdelnabi and Mario Fritz.
\newblock Adversarial watermarking transformer: Towards tracing text provenance
  with data hiding.
\newblock In {\em 42nd IEEE Symposium on Security and Privacy}, 2021.

\bibitem{abdi2010principal}
Herv{\'e} Abdi and Lynne~J Williams.
\newblock Principal component analysis.
\newblock {\em Wiley interdisciplinary reviews: computational statistics},
  2(4):433--459, 2010.

\bibitem{anderson2000null}
David~R Anderson, Kenneth~P Burnham, and William~L Thompson.
\newblock Null hypothesis testing: problems, prevalence, and an alternative.
\newblock {\em The journal of wildlife management}, pages 912--923, 2000.

\bibitem{bommasani2021opportunities}
Rishi Bommasani et~al.
\newblock On the opportunities and risks of foundation models.
\newblock {\em arXiv preprint arXiv:2108.07258}, 2021.

\bibitem{chalmers1992syntactic}
David~J Chalmers.
\newblock Syntactic transformations on distributed representations.
\newblock {\em Connectionist Natural Language Processing: Readings from
  Connection Science}, pages 46--55, 1992.

\bibitem{chen2019deepmarks}
Huili Chen, Bita~Darvish Rouhani, Cheng Fu, Jishen Zhao, and Farinaz
  Koushanfar.
\newblock Deepmarks: A secure fingerprinting framework for digital rights
  management of deep learning models.
\newblock In {\em Proceedings of the 2019 on International Conference on
  Multimedia Retrieval}, pages 105--113, 2019.

\bibitem{christ2023undetectable}
Miranda Christ, Sam Gunn, and Or~Zamir.
\newblock Undetectable watermarks for language models.
\newblock {\em arXiv preprint arXiv:2306.09194}, 2023.

\bibitem{darvish2019deepsigns}
Bita Darvish~Rouhani, Huili Chen, and Farinaz Koushanfar.
\newblock Deepsigns: An end-to-end watermarking framework for ownership
  protection of deep neural networks.
\newblock In {\em ASPLOS}, pages 485--497, 2019.

\bibitem{devlin2018bert}
Jacob Devlin, Ming-Wei Chang, Kenton Lee, and Kristina Toutanova.
\newblock Bert: Pre-training of deep bidirectional transformers for language
  understanding.
\newblock {\em arXiv preprint arXiv:1810.04805}, 2018.

\bibitem{guo-etal-2023-hc3}
Biyang Guo, Xin Zhang, Ziyuan Wang, Minqi Jiang, Jinran Nie, Yuxuan Ding,
  Jianwei Yue, and Yupeng Wu.
\newblock How close is chatgpt to human experts? comparison corpus, evaluation,
  and detection.
\newblock {\em arXiv preprint arxiv:2301.07597}, 2023.

\bibitem{he2022protecting}
Xuanli He et~al.
\newblock Protecting intellectual property of language generation apis with
  lexical watermark.
\newblock In {\em AAAI}.

\bibitem{he2021protecting}
Xuanli He, Qiongkai Xu, Lingjuan Lyu, Fangzhao Wu, and Chenguang Wang.
\newblock Protecting intellectual property of language generation apis with
  lexical watermark.
\newblock {\em arXiv preprint arXiv:2112.02701}, 2021.

\bibitem{he2022cater}
Xuanli He, Qiongkai Xu, Yi~Zeng, Lingjuan Lyu, Fangzhao Wu, Jiwei Li, and Ruoxi
  Jia.
\newblock Cater: Intellectual property protection on text generation apis via
  conditional watermarks.
\newblock {\em Advances in Neural Information Processing Systems},
  35:5431--5445, 2022.

\bibitem{jang2016categorical}
Eric Jang, Shixiang Gu, and Ben Poole.
\newblock Categorical reparameterization with gumbel-softmax.
\newblock {\em arXiv preprint arXiv:1611.01144}, 2016.

\bibitem{keskisarkka2012automatic}
Robin Keskis{\"a}rkk{\"a}.
\newblock Automatic text simplification via synonym replacement, 2012.

\bibitem{kim2003text}
Young-Won Kim, Kyung-Ae Moon, and Il-Seok Oh.
\newblock A text watermarking algorithm based on word classification and
  inter-word space statistics.
\newblock In {\em ICDAR}, pages 775--779. Citeseer, 2003.

\bibitem{kirchenbauer2023reliability}
John Kirchenbauer et~al.
\newblock On the reliability of watermarks for large language models.
\newblock {\em arXiv preprint arXiv:2306.04634}, 2023.

\bibitem{kirchenbauer2023watermark}
John Kirchenbauer, Jonas Geiping, Yuxin Wen, Jonathan Katz, Ian Miers, and Tom
  Goldstein.
\newblock A watermark for large language models.
\newblock {\em arXiv preprint arXiv:2301.10226}, 2023.

\bibitem{kuditipudi2023robust}
Rohith Kuditipudi, John Thickstun, Tatsunori Hashimoto, and Percy Liang.
\newblock Robust distortion-free watermarks for language models.
\newblock {\em arXiv preprint arXiv:2307.15593}, 2023.

\bibitem{lancaster2023artificial}
Thomas Lancaster.
\newblock Artificial intelligence, text generation tools and chatgpt--does
  digital watermarking offer a solution?
\newblock {\em International Journal for Educational Integrity}, 19(1):10,
  2023.

\bibitem{li2023protecting}
Zongjie Li, Chaozheng Wang, Shuai Wang, and Cuiyun Gao.
\newblock Protecting intellectual property of large language model-based code
  generation apis via watermarks.
\newblock In {\em CCS}, pages 2336--2350, 2023.

\bibitem{liu2023private}
Aiwei Liu, Leyi Pan, Xuming Hu, Shu'ang Li, Lijie Wen, Irwin King, and Philip~S
  Yu.
\newblock A private watermark for large language models.
\newblock {\em arXiv preprint arXiv:2307.16230}, 2023.

\bibitem{megias2022architecture}
David Meg{\i}as et~al.
\newblock Architecture of a fake news detection system combining digital
  watermarking, signal processing, and machine learning.
\newblock {\em Journal of Wireless Mobile Networks, Ubiquitous Computing, and
  Dependable Applications}, 13(1):33--55, 2022.

\bibitem{merity2016pointer}
Stephen Merity, Caiming Xiong, James Bradbury, and Richard Socher.
\newblock Pointer sentinel mixture models.
\newblock {\em arXiv preprint arXiv:1609.07843}, 2016.

\bibitem{mikolov2013efficient}
Tomas Mikolov, Kai Chen, Greg Corrado, and Jeffrey Dean.
\newblock Efficient estimation of word representations in vector space.
\newblock {\em arXiv preprint arXiv:1301.3781}, 2013.

\bibitem{mikolov2013distributed}
Tomas Mikolov, Ilya Sutskever, Kai Chen, Greg~S Corrado, and Jeff Dean.
\newblock Distributed representations of words and phrases and their
  compositionality.
\newblock {\em Advances in neural information processing systems}, 26, 2013.

\bibitem{mukherjee2023orca}
Subhabrata Mukherjee, Arindam Mitra, Ganesh Jawahar, Sahaj Agarwal, Hamid
  Palangi, and Ahmed Awadallah.
\newblock Orca: Progressive learning from complex explanation traces of gpt-4,
  2023.

\bibitem{munyer2023deeptextmark}
Travis Munyer and Xin Zhong.
\newblock Deeptextmark: Deep learning based text watermarking for detection of
  large language model generated text.
\newblock {\em arXiv preprint arXiv:2305.05773}, 2023.

\bibitem{narayan2022can}
Avanika Narayan, Ines Chami, Laurel Orr, Simran Arora, and Christopher R{\'e}.
\newblock Can foundation models wrangle your data?
\newblock {\em arXiv preprint arXiv:2205.09911}, 2022.

\bibitem{neekhara2022facesigns}
Paarth Neekhara, Shehzeen Hussain, Xinqiao Zhang, Ke~Huang, Julian McAuley, and
  Farinaz Koushanfar.
\newblock Facesigns: semi-fragile neural watermarks for media authentication
  and countering deepfakes.
\newblock {\em arXiv preprint arXiv:2204.01960}, 2022.

\bibitem{sivesind_2023}
{Nicolai Thorer Sivesind}.
\newblock Chatgpt-generated-abstracts, 2023.

\bibitem{openai2023gpt}
R~OpenAI.
\newblock Gpt-4 technical report.
\newblock {\em arXiv}, pages 2303--08774, 2023.

\bibitem{gpt-4-link}
{OpenAI Team}.
\newblock {GPT-4}.
\newblock \url{https://openai.com/research/gpt-4}.

\bibitem{orr2022data}
Laurel~J Orr, Karan Goel, and Christopher R{\'e}.
\newblock Data management opportunities for foundation models.
\newblock In {\em CIDR}, 2022.

\bibitem{papineni2002bleu}
Kishore Papineni, Salim Roukos, Todd Ward, and Wei-Jing Zhu.
\newblock Bleu: a method for automatic evaluation of machine translation.
\newblock In {\em ACL}, pages 311--318, 2002.

\bibitem{qiao2023novel}
Tong Qiao, Yuyan Ma, Ning Zheng, Hanzhou Wu, Yanli Chen, Ming Xu, and Xiangyang
  Luo.
\newblock A novel model watermarking for protecting generative adversarial
  network.
\newblock {\em Computers \& Security}, 127:103102, 2023.

\bibitem{raffel2020exploring}
Colin Raffel et~al.
\newblock Exploring the limits of transfer learning with a unified text-to-text
  transformer.
\newblock {\em The Journal of Machine Learning Research}, 21(1):5485--5551,
  2020.

\bibitem{schulman2022chatgpt}
John Schulman et~al.
\newblock Chatgpt: Optimizing language models for dialogue.
\newblock {\em OpenAI blog}, 2022.

\bibitem{tang2023science}
Ruixiang Tang, Yu-Neng Chuang, and Xia Hu.
\newblock The science of detecting llm-generated texts.
\newblock {\em Communications of the ACM}, 2024.

\bibitem{alpaca}
Rohan Taori, Ishaan Gulrajani, Tianyi Zhang, Yann Dubois, Xuechen Li, Carlos
  Guestrin, Percy Liang, and Tatsunori~B. Hashimoto.
\newblock Stanford alpaca: An instruction-following llama model.
\newblock \url{https://github.com/tatsu-lab/stanford_alpaca}, 2023.

\bibitem{wolf-etal-2020-transformers}
Hugging~Face Team.
\newblock Transformers: State-of-the-art natural language processing.
\newblock In {\em EMNLP}.

\bibitem{pytorch}
{Torch Contributors}.
\newblock {PyTorch}.
\newblock \url{https://pytorch.org/}, 2023.
\newblock {Last Access on December 26, 2022}.

\bibitem{touvron2023llama}
Hugo Touvron et~al.
\newblock Llama 2: Open foundation and fine-tuned chat models.
\newblock {\em arXiv preprint arXiv:2307.09288}, 2023.

\bibitem{wu2019machine}
Lijun Wu, Jinhua Zhu, et~al.
\newblock Machine translation with weakly paired documents.
\newblock In {\em EMNLP}, pages 4375--4384, 2019.

\bibitem{yang2023watermarking}
Xi~Yang, Kejiang Chen, Weiming Zhang, Chang Liu, Yuang Qi, Jie Zhang, Han Fang,
  and Nenghai Yu.
\newblock Watermarking text generated by black-box language models.
\newblock {\em arXiv preprint arXiv:2305.08883}, 2023.

\bibitem{yoo2023robust}
KiYoon Yoo, Wonhyuk Ahn, Jiho Jang, and Nojun Kwak.
\newblock Robust multi-bit natural language watermarking through invariant
  features.
\newblock In {\em Proceedings of the 61st Annual Meeting of the Association for
  Computational Linguistics (Volume 1: Long Papers)}, pages 2092--2115, 2023.

\bibitem{zhang2021deep}
Jie Zhang et~al.
\newblock Deep model intellectual property protection via deep watermarking.
\newblock {\em IEEE Transactions on Pattern Analysis and Machine Intelligence},
  44(8):4005--4020, 2021.

\bibitem{zhang2022opt}
Susan Zhang et~al.
\newblock Opt: Open pre-trained transformer language models, 2022.

\bibitem{zhang2019bertscore}
Tianyi Zhang, Varsha Kishore, Felix Wu, Kilian~Q Weinberger, and Yoav Artzi.
\newblock Bertscore: Evaluating text generation with bert.
\newblock {\em arXiv preprint arXiv:1904.09675}, 2019.

\bibitem{zhao2023provable}
Xuandong Zhao, Prabhanjan Ananth, Lei Li, and Yu-Xiang Wang.
\newblock Provable robust watermarking for ai-generated text.
\newblock {\em arXiv preprint arXiv:2306.17439}, 2023.

\bibitem{zhao2023protecting}
Xuandong Zhao, Yu-Xiang Wang, and Lei Li.
\newblock Protecting language generation models via invisible watermarking.
\newblock {\em arXiv preprint arXiv:2302.03162}, 2023.

\end{thebibliography}

\appendix

\clearpage
\section*{Appendix}

\section{Visualization Examples}
\label{sec:append_wm_example}
\begin{figure*}[!ht]
    \centering
    \includegraphics[width= 0.51\columnwidth]{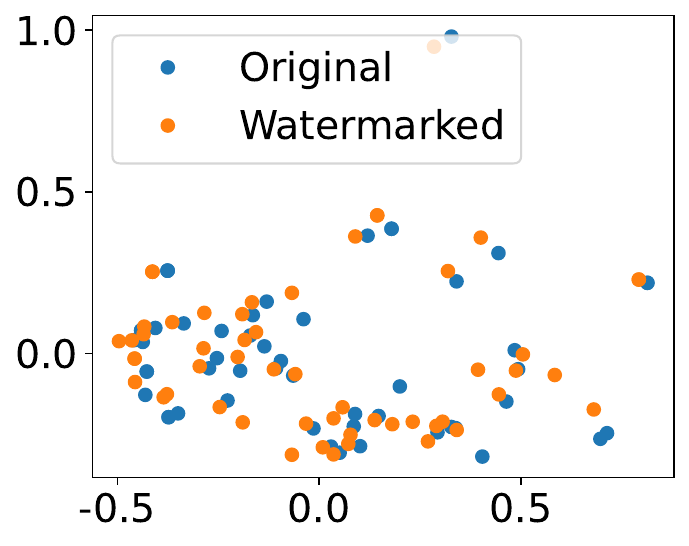}
    \includegraphics[width= 0.51\columnwidth]{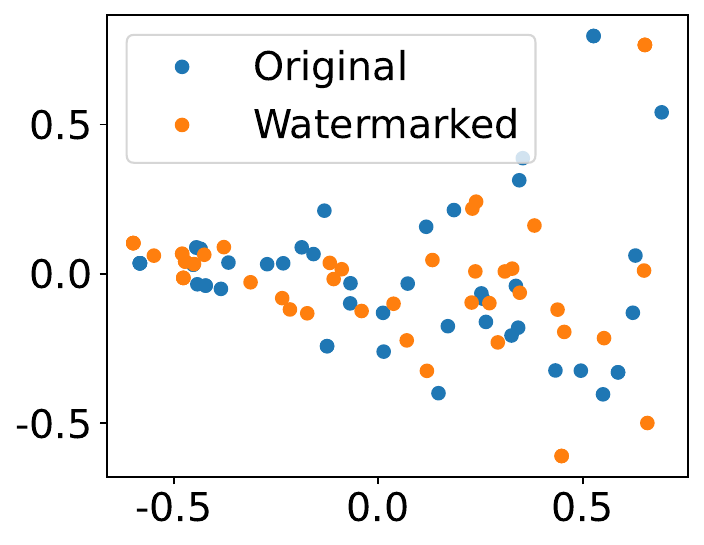}
    \includegraphics[width= 0.51\columnwidth]{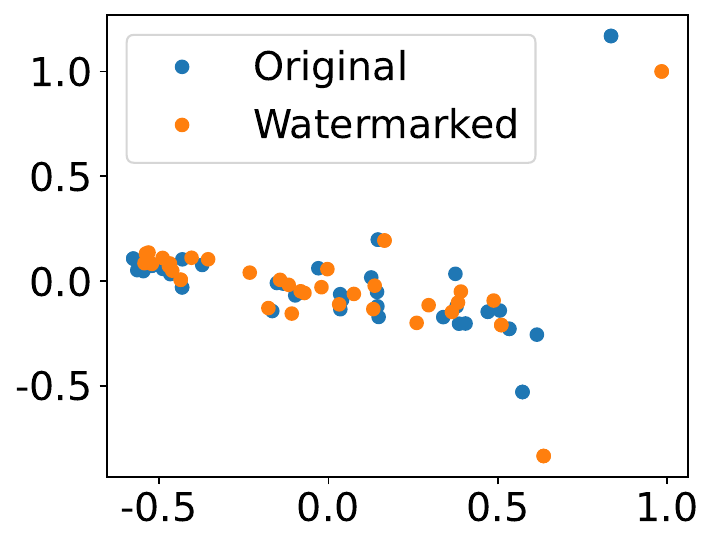}
    \includegraphics[width= 0.51\columnwidth]{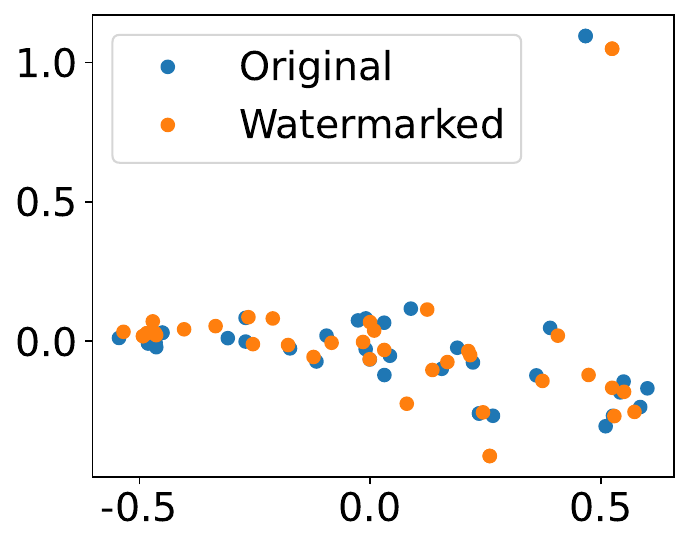} 
     \caption{The visualization of the Original LLM-generated texts and the Watermarked texts at the feature embedding level. From left to right, the figures correspond to the examples in Table~\ref{tab:visual}}
    \label{fig:visual_feat}
\end{figure*}

In this section, we include more watermarked examples generated by \sys{} from different datasets. We first show the non-wateramrked and watermark text pairs in Table~\ref{tab:append_visual}. Then, we visualize their distributions in the feature space in Figure~\ref{fig:visual_feat}. 

\textbf{More watermarked examples}
We include more \sys's watermarked examples in Table~\ref{tab:append_visual}, where all of the watermarks are successfully extracted.
From here, we can see the watermarked texts are fluent and preserve the semantics of the original texts. 

\textbf{Feature Space Visualization}
The distribution of the watermarked texts and the original LLM-generated texts at the feature embedding space is shown in Figure~\ref{fig:visual_feat}. We first obtain the texts' embeddings and reduce them to two dimensions using Principal Component Analysis (PCA)~\cite{abdi2010principal}. Those reduced dimensions are plotted in Figure~\ref{fig:visual_feat}. From the figure, we can find the watermarked texts and LLM-generated texts are close to each other in the feature embedding space. This ensures the watermarked texts do not alter the meaning of the original texts. Their distribution patterns are different, which is the basis for the message decoding module to learn how to predict message bits from the watermarked texts. 

\section{Experimental Setups}~\label{append_setup}
In this section, we first introduce the infrastructure for training \sys{}. Then, we include \sys's detailed architecture configurations and the \sys's hyperparameters at both training and inference time.  

\textbf{Hardware Infrastructure} Our code is implemented using PyTorch~\cite{pytorch} version 1.9.0 and open-source transformer models from huggingface~\cite{wolf-etal-2020-transformers}. The training and inference of our watermarking models are performed on NVIDIA RTX A6000 GPUs with Ubuntu 20.04.5 LTS and AMD Ryzen Threadripper 3990X 64-Core Processors.

\textbf{\sys{} Architecture} ~\label{sec:parameter_define}
We include \sys{} architecture details in Table~\ref{tab:archi}. Our T5 model~\cite{raffel2020exploring} is initialized using the official pre-trained checkpoints from huggingface~\cite{wolf-etal-2020-transformers}. The rest of the submodule models are trained from scratch. For the transformer-based models, $n\_head$ is the number of heads for attention modules, $N$ is the number of layers, and $d\_model$ is the feedforward dimension.

\begin{table}[!ht]
  \centering
  \small
  \begin{tabular}{lccc}
    \toprule
   SubModule  & Backbones & Input Size & Output Size\\
    \midrule
     Message $\mathbf{R}_m$ & Linear & ML & 512 \\
     Mapping $\mathbf{R}_e$ & Linear & TL & ED \\
    \bottomrule  
  \end{tabular}
  \begin{tabular}{lcccc}
  \toprule
   SubModule  & Backbones & $n\_head$ & $N$ & $d\_model$\\
    \midrule
   Seq2Seq $\mathbf{S}_e$ & T5 Encoder& 12& 12& 768\\
   Seq2Seq $\mathbf{S}_d$ & T5 Decoder& 12& 12& 768\\
   Extractor $\mathbf{E}$  & Transformer & 8 & 3 & 512 \\
    \bottomrule  
    \end{tabular}
  \caption{\sys{} architecture details. ML is short for Message Length, TL is short for Token Length, ED is short for Tokenizer Embedding Dimension \label{tab:archi}}
  \vspace{-10pt}
\end{table}

\begin{figure}[!ht]
    \centering
    \includegraphics[width= 0.49\columnwidth]{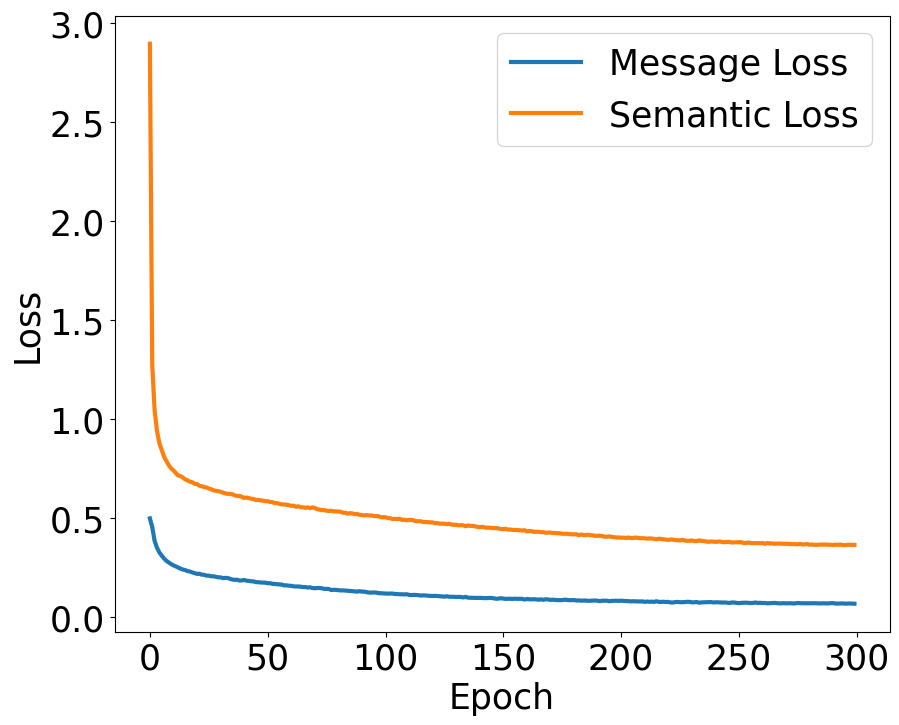}
    \includegraphics[width= 0.49\columnwidth]{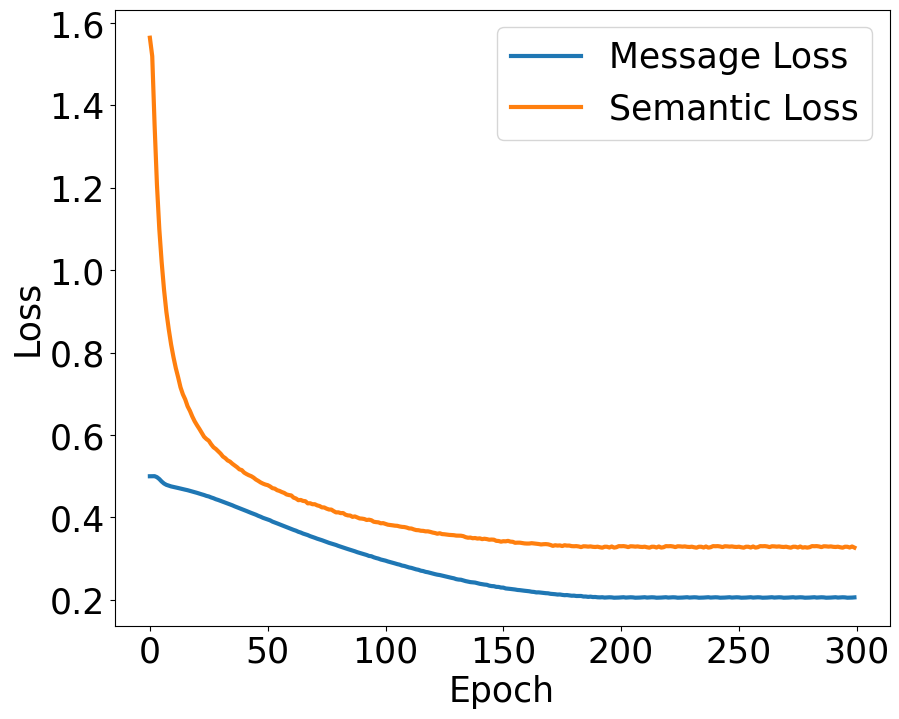}
     \caption{The loss curve for training \sys{} on the HC3 dataset at the segment level and inserting 8-bit and 16-bit messages, respectively. The message recovery loss is $L_M$ and the semantic loss is $L_S$.}
    \label{fig:loss}
     \vspace{-10pt}
\end{figure}

\textbf{\sys{} Training Details}
We add additional training details and the hyperparameters in Table~\ref{tab:hyper3}.  
The watermark message is encoded vis the L1 loss between the input and extracted message from the Message Decoding module. The watermarked texts ensure their semantic similarity by minimizing the cross-entropy between the original texts and the watermarked distribution from the Message Encoding module. For the text transformation, we set 33\% possibility for $\mathcal{S}_t(T+M)$ to be Text Replacement Attacks, 33\% possibility to be Text Deletion Attacks, and 34\% possibility to be Text Addition Attacks. The loss curve in Figure~\ref{fig:loss} is the plot for training \sys{} on the ChatGPT Abstract Dataset and inserting different message bits. We can find both the message recovery loss and the semantic loss converge at the end of the training.

\textbf{Watermark Insertion and Extraction}
For the optimized beam search algorithm in the inference time, the detailed hyperparameters for watermark insertion are in Table~\ref{tab:hyper3}. We choose the texts with the highest watermark extraction accuracy to be the watermarked texts for output. For the watermark extraction, we use the trained $\mathbf{R}_e$ to map the one-hot token distribution to the embedding space and extract the watermarks via the extractor $\mathbf{E}$. 
 %\begin{table}[!ht]
 %  \centering
 %  \small
 %  \begin{tabular}{lc}
 %    \toprule
 %  Training-time  Parameters & Settings\\
 %    \midrule
 %    Epoch, Batch size & 300, 16 \\
 %   $w_w$, $w_t$ $w_1$, $w_2$ & 0.7, 0.3, 1, 1\\
 %    Maximum Token Size &  80\\
  %   Optimizer, Learning rate  & AdamW, 3e-5 \\
  %   Gumbel Temperature & 0.3\\
  %   Mask Percentage & 50\% \\
  %   \bottomrule  
  %   \toprule
  %  Inference-time Parameters & Settings\\
  %   \midrule
  %   Beam Width, Repeat & 5, 5 \\
  %   Gumbel Temperature & \{1, 1, 1.5, 1.5, 2\}\\
  %   Mask Percentage & 50\% \\
  %   \bottomrule  
  % \end{tabular}
   %\caption{\sys{} hyperparameters in both training and inference time.\label{tab:hyper}}
 %\end{table}

\begin{figure}[h]
    \begin{minipage}[t]{0.49\linewidth}
    \vspace{10pt}
   \centering 
    \resizebox{\textwidth}{!}{
    \setlength\tabcolsep{3pt}
   \begin{tabular}{lc}
    \toprule
  Training-time & Settings\\
    \midrule
    Epoch, Batch size & 300, 16 \\
   $w_w$, $w_t$ $w_1$, $w_2$ & 0.7, 0.3, 1, 1\\
    Maximum Token Size &  80\\
    Optimizer, Learning rate  & AdamW, 3e-5 \\
    Gumbel Temperature & 0.3\\
   Mask Percentage & 50\% \\
    \bottomrule  
  \end{tabular}}
    \end{minipage} 
    \hfill
     \begin{minipage}[t]{0.5\linewidth}
    \vspace{10pt}
    \centering 
    \resizebox{\textwidth}{!}{
    \setlength\tabcolsep{3pt}
   \begin{tabular}{lc}
    \toprule
   Inference-time & Settings\\
    \midrule
   Beam Width, Repeat & 5, 5 \\
   Gumbel Temperature & \{1, 1, 1.5, 1.5, 2\}\\
    Mask Percentage & 50\% \\
    \bottomrule  
\end{tabular}}
    
    \end{minipage} 
   \captionof{table}{\sys{} Hyperparameters during training/inference.}\label{tab:hyper3}
     \vspace{-0.5cm}
\end{figure}

\begin{table*}[!ht]
\small
\centering
\resizebox{\textwidth}{!}{%
\begin{tabular}{p{10cm}p{10cm}}
\toprule
Original Text &  Watermarked Text  \\ \midrule
This means that you \hl{may} not be able to buy \hl{or} sell stocks \hl{or} other investments, pay bills, or make \hl{other financial} transactions. However, there \hl{may} be some \hl{limited} options \hl{for you} to \hl{handle} your finances. &This means that you \hl{simply} not be able to buy \hl{and} sell stocks \hl{,} other investments, pay bills, or make \hl{any significant} transactions. However, there \hl{might} be some \hl{other} options \hl{available } to \hl{manage} your finances.\\\hline 
LinkedIn is a social networking \hl{platform} that is \hl{primarily} used \hl{for professional} networking. It is a place where people can \hl{create} a profile, connect with other professionals, and share their professional experiences and \hl{skills}. \hl{LinkedIn} is often used by job seekers to find employment \hl{opportunities} and by \hl{employers} to \hl{find} and recruit qualified candidates for job \hl{openings}. \hl{LinkedIn} can also be used to connect with industry \hl{experts}.
& LinkedIn is a social networking \hl{website} that is \hl{widely} used \hl{by business} networking. It is a place where people can \hl{build} a profile, connect with other professionals, and share their professional experiences and \hl{knowledge}. \hl{It} is often used by job seekers to find employment\hl{,} and by \hl{businessman} to \hl{search} and recruit qualified candidates for job \hl{postings}. \hl{It} can also be used to connect with industry \hl{professionals}.\\\hline 
Hedge funds are investment \hl{funds} that use a variety of strategies \hl{to} try to generate higher returns \hl{for} their investors. They are usually started \hl{by} a group of investment professionals who have experience in the financial \hl{industry and} want to start \hl{their} own business. To start a hedge fund, the founders typically need to \hl{raise} money from investors, such as wealthy individuals or \hl{institutional} investors like pension funds.
& Hedge funds are investment \hl{companies} that use a variety of strategies \hl{inside} try to generate higher returns \hl{on} their investors. They are usually started \hl{when} a group of investors professionals who have experience in the financial \hl{sector directly} want to start own business. To start a hedge fund, the founders typically need to \hl{have} money from investors, such as wealthy individuals or \hl{companies} investors like pension funds.\\\hline 
WEP, WPA, and WPA2 are different types of security protocols \hl{that} can be used to \hl{protect a} wireless network. Here's a \hl{breakdown} of \hl{the} main \hl{advantages} and disadvantages of each: WEP (\hl{Wired} Equivalent Privacy): Advantages: WEP was one of the first security protocols. & WEP, WPA and WPA2 are different types of security protocols can be used to \hl{secure} wireless network. Here's a \hl{simple} of \hl{some} main \hl{differences} and disadvantages of each: WEP (\hl{Wire} Equivalent Privacy): Advantages:  WEP was one of the first security protocols. \\\hline  
In this paper, we \hl{investigate} the behavior \hl{of} the Goldstone \hl{modes and} the Higgs condensation beyond the one-loop approximation in the context of the \hl{Standard} Model of particle physics. We show that, \hl{even though} the one-loop calculation \hl{provides} a reasonable \hl{description} of the phenomenon in most cases.
& In this paper, we \hl{discuss} the behavior the Goldstone \hl{project in} the Higgs condensation beyond the one-loop approximation in the context of \hl{super dynamics} Model of particle physics. We show that,  \hl{at} the one-loop \hl{represents} a reasonable \hl{explanation} of the phenomenon in most cases. \\\hline  
\hl{A} version of Sonic the Hedgehog was \hl{developed} by Ancient \hl{and released} in 1991 for Sega's \hl{8 @-@} bit consoles, the \hl{Master System and} Game. \hl{Its } plot and \hl{gameplay} mechanics \hl{are} similar to the 16 @-@ bit \hl{version}, \hl{with} different level \hl{themes} and digital assets.  
& \hl{The} version of Sonic the Hedgehog was \hl{released} by Ancient \hl{Come} in 1991 for Sega's \hl{16- (( bit maps}, the \hl{original} Game. \hl{The} plot and \hl{the} mechanics \hl{alone} similar to the 16-- bit \hl{console} — \hl{but} different level \hl{settings} and digital assets. \\\hline
The \hl{artist} you are thinking about is \hl{Salvador Dal}, a prominent Spanish artist who is best known for his \hl{unique style} and \hl{surrealist paintings}. Three of his most \hl{recognized} works \hl{are}: 1. The \hl{Persistence} of Memory: Also known as "Soft Watches," this painting was created in \hl{1931}. It features melting clocks.
& The \hl{one} you are thinking of is \hl{William Dali}, a prominent Spanish artist who is best known for his \hl{complex} and \hl{surreist elements}. Three of his most \hl{famous} works \hl{include}: 1. The \hl{Pers} of Memory: Also known as "Soft Watches" this painting was created in \hl{1913}. It features melting clocks.\\\hline
Here are four \hl{popular} social media platforms: \hl{1.} Facebook: Facebook is a popular platform for connecting with friends, family, and other \hl{acquaintances}. Users can share updates, \hl{photos}, and videos, and \hl{engage} with others through \hl{comments}, likes, and shares. \hl{2.} Instagram: Instagram is a photo and \hl{video}-sharing \hl{platform} where users can share their stories and \hl{daily} life.
& Here are four \hl{common} social media platforms: Facebook: Facebook is a popular platform for connecting with friends, family, and other \hl{connecteds}. Users can share updates, \hl{tips}, and videos, and \hl{connect} with others through \hl{posts}, likes, and shares. Instagram: Instagram is a photo \hl{sharing} and \hl{picture}-sharing \hl{website} where users can share their stories and \hl{everyday} life.\\\hline
A neural network \hl{is} an artificial intelligence (AI) \hl{model inspired} by the human \hl{brain} and its \hl{structure} and \hl{functionality}. It is designed to process \hl{information} and make \hl{predictions} or \hl{decisions} based on \hl{the} input data. It consists of multiple layers of \hl{neurons} (\hl{or} nodes), which \hl{are connected} by weights and biases. 
& A neural network\hl{,} an artificial intelligence (AI) \hl{system motivated} by the human \hl{mind} and its \hl{complex} and \hl{function}. It is designed to process \hl{data} and make \hl{decisions} or \hl{predictions} based on input data. It consists of multiple layers of \hl{layers} (\hl{called} nodes), which \hl{perform assigned} by weights and biases.\\\hline
Create a question about the consequences of global warming. \hl{The question could be about}: - The impact of global warming on \hl{individual} countries or regions - The effects of global warming on animal and plant \hl{species} - The \hl{influence} of global warming on human \hl{societies and economies} - The role of global warming in \hl{exacerbating} natural disasters.
& Create a question about the consequences of global warming. \hl{The following might include}: "- the impact of global warming on \hl{different} countries or regions - The effects of global warming on animal and plant \hl{populations} - The \hl{role} of global warming on human \hl{health and societies} - The role of global warming in \hl{exaggerating} natural disasters ".\\
\bottomrule
\end{tabular}}
\caption{More  Watermarked Text Examples from \sys. The edited words are highlighted in \hl{yellow}. The first six examples are randomly taken from ChatGPT abstract, Human Abstract, HC3, and WikiText-2 datasets. The last three are randomly taken from Alpaca Datasets instructed LLaMA-2 generated texts. 
\label{tab:append_visual}}
\end{table*}

%%%%%%%%%%%%%%%%%%%%%%%%%%%%%%%%%%%%%%%%%%%%%%%%%%%%%%%%%%%%%%%%%%%%%%%%%%%%%%%%
\end{document}